 \documentclass[prb,aps,twocolumn]{revtex4} 
\usepackage{graphicx} 
\usepackage{tabularx}
\usepackage{dcolumn} 
\usepackage{color}
\usepackage{bm}
\usepackage{amsmath}
\usepackage{amssymb}

\begin{document} 


\title{Non-unitary triplet superconductivity tuned by field-controlled magnetization --URhGe, UCoGe and UTe$_2$--
}

\author{Kazushige Machida} 
\affiliation{Department of Physics, Ritsumeikan University, 
Kusatsu 525-8577, Japan} 

\date{\today}

\begin{abstract}
We report on a theoretical study on ferromagnetic
superconductors, URhGe, and UCoGe and identify the pairing state as a non-unitary
spin-triplet one with time reversal symmetry broken, analogous to superfluid $^3$He-A phase. A recently found
superconductor UTe$_2$ with almost ferromagnet is analyzed by the same manner.
Through investigating their peculiar upper critical field $H_{\rm c2}$ shapes,
it is shown that the pairing symmetry realized in all three compounds can be tuned by their
magnetization curves under applied fields. This leads to the reentrant $H_{\rm c2}$ in URhGe,
an S-shaped in UCoGe and an L-shaped $H_{\rm c2}$ in UTe$_2$ observed under the field direction parallel to the
magnetic hard $b$-axis in orthorhombic crystals in common.
The identification with double chiral form: ${\bf d}(k)=(\hat{b}+i\hat{c})(k_b+ik_c)$ in UTe$_2$ naturally 
enables us to understand (1) multiple phases with A$_1$, A$_2$, and A$_0$ phases observed under pressure, (2)
the enhanced reentrant $H_{\rm c2}$ for the off-axis
direction is associated with first order meta-magnetic transition, and (3)
Weyl point nodes oriented along the magnetic easy $a$-axis. All three compounds 
 are found to be topologically rich solid-state materials worth further investigation. 
 \end{abstract}

\maketitle 

\section{Introduction}

The competing orders are at the heart of the strongly correlated systems in general
where multiple long-range or short range orderings, such as superconductivity (SC), ferromagnetism (FM),
spin and charge density waves are emerging out of the strong interactions in condensed matter systems.
This is particularly true in the case of unconventional superconductivity, which is often associated with
other orderings mentioned~\cite{kato1,kato2}. A good example is high temperature cuprate superconductors in which
various coexisting or mutually repulsive orderings are found~\cite{kivelson}.

There has been much attention focused on 
ferromagnetic superconductors~\cite{aokireview}, such as 
UGe$_2$~\cite{UGe2}, URhGe~\cite{URhGe}, and UCoGe~\cite{UCoGe} in recent years.
A new member of such a superconductor, UTe$_2$ with $T_{\rm c}$=1.6K~\cite{ran,aoki2}, 
which is almost ferromagnetic is discovered quite recently and attracts much excitement. 
Those systems are contrasted with the coexisting materials of magnetism and superconductivity in
(RE)Rh$_4$B$_4$ (RE: 4f rare earth elements) and Chevrel compounds (RE)Mo$_6$S$_8$ in 
that the 4f electrons responsible for magnetism
are localized spatially and distinctive from the conduction electrons~\cite{machidareview}.
Here the 5f electrons responsible for magnetism are more subtle in that they participate both magnetism and superconductivity.

UTe$_2$ has been investigated experimentally~\cite{knebel,daniel,miyake,ran2,metz,mad,
tokunaga,sonier,nakamine,hayes,1,2,3,4,rosa,aokiP,andreyUTe2,kittaka} and 
theoretically~\cite{xu,ishizuka,shick,nevi,fidrysiak,yarzhemsky,fujimoto,lebed,kmiyake}.
Simultaneously renewed interest on the former three compounds are developing.
These heavy Fermion materials belong to a strongly correlated system
that is heavily governed by the 5f electrons, which form a coherent narrow band with a 
large mass enhancement below the Kondo temperature.
Because the upper critical field $H_{\rm c2}$ in those compounds exceeds the Pauli paramagnetic limitation,
a triplet or odd parity pairing state is expected to be realized~\cite{aokireview}.
However, detailed studies of the pairing symmetry remain lacking despite of the fact that 
previous knowledge of the first three compounds is accumulated
for over two decades.
Thus now it is a good chance to understand those ``old'' materials URhGe and UCoGe together with the new UTe$_2$
by seeking some common features. 

The prominent SC properties observed commonly in these superconductors are as follows:
When $H$ is applied parallel to the magnetic hard $b$-axis  in orthorhombic crystals, $H_{\rm c2}$ exhibits
the reentrant behavior in URhGe, where the SC state that disappeared reappears
at higher fields~\cite{levy}, or an S-shape in UCoGe~\cite{aokiS} and an L-shape in UTe$_2$~\cite{knebel} 
in the $H$-$T$ plane.
Above the superconducting transition temperature $T_{\rm c}$, ferromagnetic transition occurs 
in URhGe and UCoGe. Thus, the SC state survives under a strong internal field, resulting from an exchange interaction
between the conduction and the 5f electrons. 
However, in UTe$_2$ ``static'' FM has not been detected
although FM fluctuations are probed~\cite{ran,miyake,tokunaga,sonier} above  $T_{\rm c}$, i.e.
there is a diverging static susceptibility along the magnetic easy $a$-axis~\cite{ran,miyake} and the 
nuclear relaxation time $1/T_2$  in NMR~\cite{tokunaga}. 

The gap structure is unconventional, characterized by either a point in UTe$_2$~\cite{aoki2,ran} or line 
nodes in the others~\cite{aokireview}.
There is clear experimental evidence for double transitions:
the two successive second order SC phase transitions seen 
in specific heat experiments as distinctive jumps systematically change 
under pressure ($P$) in UTe$_2$~\cite{daniel}.
A similar indication for double SC transitions in ambient pressure is 
observed in UCoGe at $T_{\rm c2}\sim$0.2K~\cite{manago1,manago2}
where the nuclear relaxation time $1/T_1T$ in NMR experiments exhibits a plateau corresponding 
to the ``half residual density of states (DOS)'' value at the intermediate $T$ below $T_{\rm c}=$0.5K.
Upon further lowering of $T$, it stars decreasing again at 0.2 K.
Recent specific heat $C/T$ data for several high quality samples of UTe$_2$~\cite{aoki2,ran,kittaka} 
commonly show the residual DOS amounting to $0.5N(0)$, which is half of the 
normal DOS $N(0)$, while some exhibit zero residual DOS~\cite{metz}. 
Thus, this ``residual'' half DOS issue is currently controversial.
We propose a method to resolve this issue, discussed later in this paper.

To understand these three spin-polarized superconductors URhGe, UCoGe, and UTe$_2$ in a unified way, 
we  develop a phenomenological theory based on
the assumption that the three compounds are coherently described 
in terms of the triplet pairing symmetry analogous to the superfluid 
$^3$He-A-phase~\cite{leggett}.
It is instructive to remember that the A$_1$-A$_2$ phase transition is induced by an applied field, 
which is observed as the clear double specific heat jumps~\cite{halperin}.
The originally degenerate transition temperatures for the A phase are split into
the A$_1$ and A$_2$ phases under applied fields~\cite{mermin}.

Therefore, to address the experimental facts mentioned above, 
we postulate the A-phase-like triplet pair symmetry, which responds to the spontaneous 
FM and/or induced moment under perpendicular external fields, to yield the A$_1$-A$_2$ double transitions.
This scenario coherently explains the observed reentrant $H_{\rm c2}$ in URhGe, the S-shape in UCoGe, and the L-shape in UTe$_2$
for the field direction along the magnetic hard $b$-axis in a unified way.

As mentioned above, the A$_1$-A$_2$ phase transition in $^3$He A-phase~\cite{halperin} is controlled by the linear 
Zeeman effect due to the applied field, which acts to split $T_{\rm c}$~\cite{mermin}. 
In the spin polarized superconductors, $T_{\rm c}$ is controlled by 
the spontaneous and/or field-induced magnetic moment, which is linearly coupled to the non-unitary order parameter. 
We employ the Ginzburg-Landau (GL) theory to describe these characteristic $H_{\rm c2}$ curves.
We also identify the pairing symmetry by group theoretic classification~\cite{machida}
based on our previous method~\cite{ozaki1,ozaki2}. The 
pairing symmetry is a non-unitary triplet~\cite{machida,ohmi,machida2}, 
where the {\bf d}-vector is a complex function that points perpendicular to
the magnetic easy axis in zero-field. The gap function possesses either a point or line node with a possibly chiral $p$-wave orbital form. 
This is maximally consistent with the SC characteristics obtained so far in UTe$_2$, such as the STM observation~\cite{mad} of chiral edge states, the polar Kerr experiment~\cite{hayes}, which shows time reversal symmetry breaking, and other various thermodynamic measurements.

The arrangement of this paper is following.
We set up the theoretical framework to explain those experimental facts in the three compounds,
URhGe, UCoGe and UTe$_2$ in Section II. The theory is based on the Ginzburg-Landau theory for the order parameter with three components.
The quasi-particle spectra in the triplet states are examined to understand thermodynamic behaviors for the materials.
In Section III we investigate the generic phase transitions of the present pairing state under fields
applied to various field  directions relative to the orthorhombic crystalline axes.
In order to prepare analyzing the experimental data for URhGe, UCoGe and UTe$_2$ which exhibit a variety of the
$H_{\rm c2}$ such as reentrant SC (RSC), S-shaped, and L-shaped one, the magnetization curves for three compounds
are studied in detail and evaluated the curves when the experimental data are absent in Section IV.
We apply the present theory to the three compounds and explain the peculiar $H_{\rm c2}$ curves observed in
Section V, including the multiple phase diagrams in UTe$_2$ under pressure.
Section VI devotes to detailed discussions on the gap structure, and pairing symmetries for each 
material. Summary and conclusion are given in the final section VII.
The present paper is a full paper version of the two short papers by the author~\cite{short1,short2}.

\section{Theoretical Framework}
\subsection{Ginzburg-Landau theory}

In order to understand a variety of experimental phenomena exhibited by the three compounds in a common theoretical framework,
we start with the most generic Ginzburg-Landau (GL) theory for a spin triplet state. This is general enough to allow us to
describe the diversity of those systems.   Among abundant spin triplet, or odd-parity paring states we assume 
an A-phase like pairing state described by the complex $\bf d$-vector with three components:

\begin{eqnarray}
{\bf d}(k)=\phi(k){\vec \eta}=\phi(k)({\vec \eta}'+i{\vec \eta}'').
\label{d-vector}
\end{eqnarray}

\noindent
${\vec \eta}'$ and ${\vec \eta}''$ are real three dimensional vectors in the spin space for Cooper pairs, and 
$\phi(k)$ is the orbital part of the pairing function. 
This is classified group-theoretically under the overall symmetry:

\begin{eqnarray}
\rm {SO}(3)_{\rm spin}\times {\rm D}^{\rm orbital}_{\rm 2h}\times {\rm U}(1)
\label{symmetry}
\end{eqnarray}

\noindent 
with the spin, orbital and gauge symmetry respectively~\cite{machida,annett}.

In this study, we adopt the weak spin-orbit coupling scheme~\cite{ozaki1,ozaki2} which covers 
the strong spin-orbit (SO)
case as a limit. The strength of the SO coupling depends on materials and
is to be appropriately tuned relative to the experimental situations.
It  will turn out to be crucial to choose the weak SO coupling case in understanding the $H_{\rm c2}$ phase diagrams with peculiar shapes: 
This allows the $\bf d$-vector rotation
under an applied field whose strength is determined by the SO coupling.
Note that in the strong SO coupling scheme the $\bf d$-vector rotation field is infinite
because the Cooper pair spin is locked to crystal lattices.

There exists U(1)$\times$Z$_2$ symmetry in this pairing, i.e., invariance
under ${\bf d}\rightarrow -{\bf d}$ and gauge transformations.
We emphasize here that this SO(3) triple spin symmetry of the pairing function is expressed by a complex
three component vectorial order parameter $\vec{\eta}=(\eta_a,\eta_b,\eta_c)$ in the most general.
It will turn out later to be important also to describe complex multiple phase diagram, consisting
of five distinctive phases, but this is a minimal framework which is necessary and sufficient.

Under the overall symmetry expressed by Eq.~(\ref{symmetry})
the most general Ginzburg-Landau free energy functional 
up to the quadratic order is written down as

\begin{eqnarray}
F^{(2)}=\alpha_0(T-T_{\rm c0}){\vec \eta}\cdot{\vec \eta}^{\star}+b|{\vec M}\cdot{\vec \eta}|^2+
cM^2{\vec \eta}\cdot{\vec \eta}^{\star}\nonumber \\
+i\kappa {\vec M}\cdot {\vec \eta}\times {\vec \eta}^{\star}
\label{f2}
\end{eqnarray}

\noindent
with $b$ and $c$ positive constants. 
The last invariant with the coefficient $\kappa$ comes from the non-unitarity of the pairing function in the presence of the 
spontaneous moment  and field induced $\vec M(H)$, which are to break the SO(3) spin space symmetry in Eq.~(\ref{symmetry}). 
We take $\kappa>0$ without loss of generality.
This term responds to external field directions differently through their magnetization curves.

The fourth order term in the GL functional is given by~\cite{machida,annett}

\begin{eqnarray}
F^{(4)}={\beta_1\over2}({\vec \eta}\cdot{\vec \eta}^{\star})^2
+{\beta_2\over2}|{\vec \eta}^2|^2 .
\label{f4}
\end{eqnarray}

\noindent
Because the fourth order terms are written as 

\begin{eqnarray}
F^{(4)}={\beta_1\over2}({\vec \eta}'\cdot{\vec \eta}'+{\vec \eta}''\cdot{\vec \eta}'')^2
+{\beta_2\over2}[({\vec \eta}'\cdot{\vec \eta}'-{\vec \eta}''\cdot{\vec \eta}'')^2 \nonumber \\
+4({\vec \eta}'\cdot{\vec \eta}'')^2]
\label{f4-2}
\end{eqnarray}

\noindent
for $\beta_1, \beta_2>0$, we can find a minimum when $|{\vec \eta}'|=|{\vec \eta}''|$
and ${\vec \eta}'\perp{\vec \eta}''$.
Notably, the weak coupling estimate~\cite{machida} leads to 
${\beta_1/\beta_2}=-2$. Thus we have to resort to the strong coupling effects in the following arguments
in order to stabilize an A$_1$ phase.

It is convenient to introduce 

\begin{eqnarray}
\eta_{\pm}={1\over \sqrt2}(\eta_b\pm i\eta_c)
\label{eta+-}
\end{eqnarray}

\noindent
for ${\bf M}=(M_a,0,0)$ where we denote the $a$-axis as the magnetic easy axis in this and next sections. From Eq.~(\ref{f2}) the quadratic term $F^{(2)}$ is rewritten in terms of $\eta_{\pm}$ and $\eta_a$ as


\begin{eqnarray}
F^{(2)}=\alpha_0\{(T-T_{\rm c1})|\eta_{+}|^2+(T-T_{\rm c2})|\eta_{-}|^2\nonumber \\
+(T-T_{\rm c3})|\eta_{a}|^2\}
\label{f2-2}
\end{eqnarray}

\noindent
with 

\begin{eqnarray}
T_{\rm c 1,2}=T_{\rm c0} \pm{\kappa\over \alpha_0}M_a, \nonumber \\
T_{\rm c 3}=T_{\rm c0} -{b\over \alpha_0}M^2_a.
\label{tc}
\end{eqnarray}

\noindent
The actual second transition temperature is modified to

\begin{eqnarray}
T^{\prime}_{\rm c2}=T_{\rm c0}-{\kappa M_a \over \alpha_0} \cdot{{\beta_1-\beta_2}\over{2\beta_2}}
\label{tc2}
\end{eqnarray}

\noindent
because of the fourth order GL terms in Eq.~(\ref{f4}).
And also $T_{\rm c 3}$ starts decreasing in the linear  $|M_a|$ in stead of $M^2_a$ mentioned above just near $|M_a|\ll 1$.
This comes from the renormalization of $T_{c3}$ in the presence of 
$|\eta_{+}|^2\propto (T_{c1}-T)$ and $|\eta_{-}|^2\propto (T_{c2}-T)$.
Those terms give rise to the $|M_a|$-linear suppression of $T_{c3}$ through fourth order terms.
Here we note that among the GL fourth order terms, $Re(\eta_a^2\eta_{+}\eta_{-})$ in Eq.~(\ref{f4}) becomes important
in interpreting the $H_{\rm c2}$ data later because it is independent of the signs of the GL parameters $\beta_1$ and $\beta_2$.
For $1\le{\beta_1/\beta_2}\le3$, 

\begin{eqnarray}
T^{\prime}_{c2}>T_{c2}=T_{\rm c0} \pm{\kappa\over \alpha_0}M_a.
\label{tc2-2}
\end{eqnarray}

\noindent
This could lead to the modification of the otherwise symmetric phase diagram:

\begin{eqnarray}
T_{c1}-T_{c0}=T_{c0}-T_{c2}.
\label{asymmetry}
\end{eqnarray}

\noindent
The fourth order contribution of Eq.~(\ref{tc2}) to $T_{c2}$ may become important to quantitatively reproduce 
the $H$-$T$ phase diagram, such as the asymmetric L-shape $H_{\rm c 2}^b$ observed in UTe$_2$~\cite{knebel}.

Note that the ratio of the specific heat jumps to

\begin{eqnarray}
{\Delta C(T_{c1})\over \Delta C(T_{c2})}={T_{c1}\over 
T_{c2}}\cdot {\beta_1\over{\beta_1+\beta_2}}.
\label{C-jump}
\end{eqnarray}

\noindent
The jump at $T_{c2}$ can be quite small for $T_{c1}\gg T_{c2}$.

\begin{figure}
\includegraphics[width=6cm]{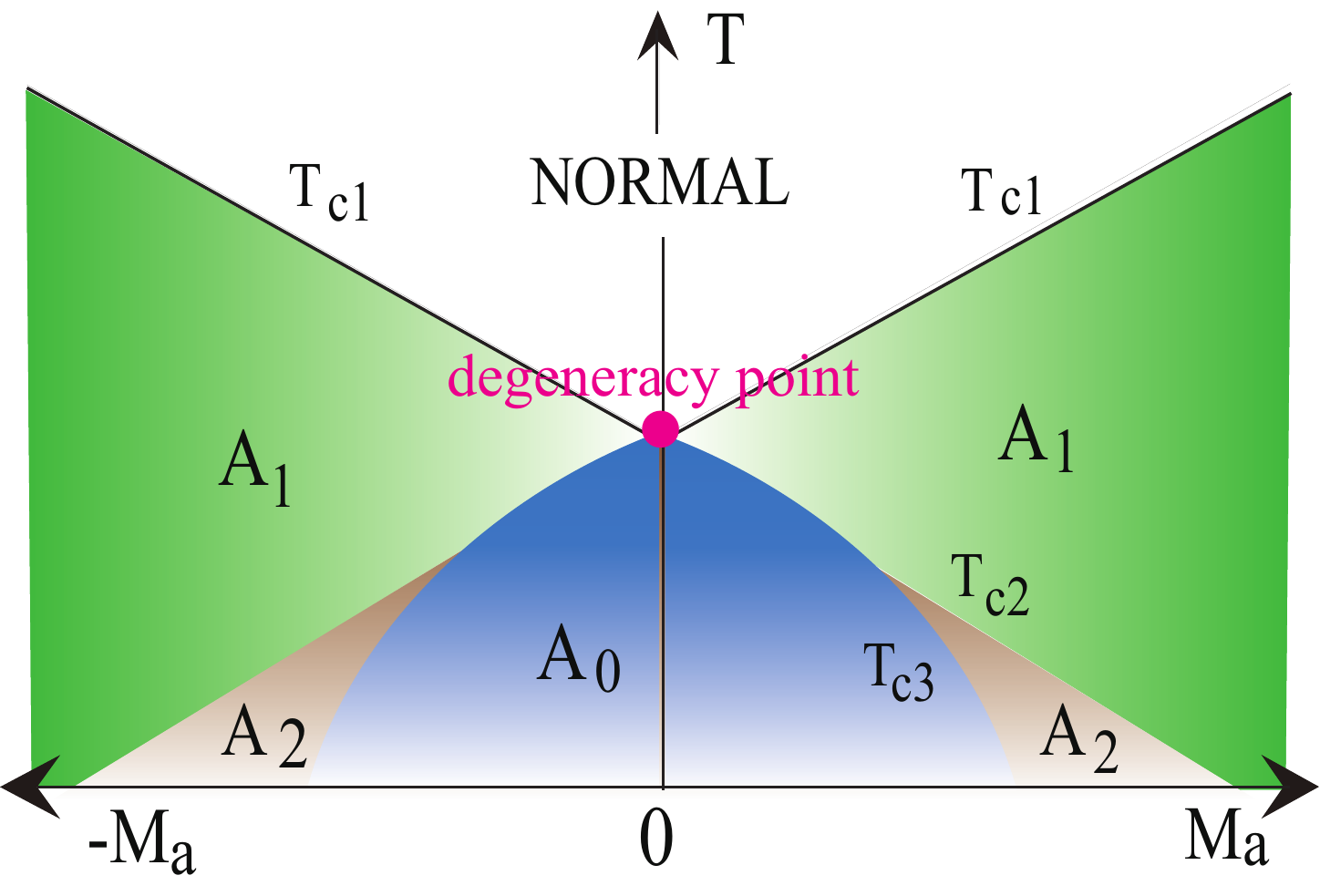}
\caption{(color online) 
Generic phase diagram in the $T$ and $M_a$ plane.
$T_{\rm c1}$ ($T_{\rm c2}$) for the A$_1$ (A$_2$) phase  increases (decreases) linearly in $M_a$.
The third phase A$_0$ decreases quadratically in $M_a$ away from the degenerate point at $M_a$=0.
}
\label{ProtoA0}
\end{figure}

The FM moment $M_a$ acts to shift the original transition temperature $T_{\rm c0}$ and split it into $T_{c1}$,
$T_{c2}$, and $T_{c3}$ as shown in Fig.~\ref{ProtoA0}.
Here, the A$_1$ and A$_2$ phases correspond to $|\uparrow\uparrow>$ pair and $|\downarrow\downarrow>$ pair, respectively and
the A$_0$ phase is $|\uparrow\downarrow>+|\downarrow\uparrow>$ for the spin quantization axis parallel to
the magnetization direction $M_a$.
According to Eq.~(\ref{tc}), $T_{c1}$ ($T_{c2}$) increases (decrease) linearly as a function of $M_a$
while $T_{c3}$ decreases quadratically as $M^2_a$ far away from the degeneracy point shown there (the red dot).
The three transition lines meet at $M_a$=0 where the 
three components $\eta_i$ ($i=+,-,a$) are all degenerate, restoring SO(3) spin space symmetry. 
Thus away from the degenerate point at $M_a$=0, the A$_0$
phase starts at $T_{\rm c3}$ quickly disappears from the phase diagram.
Below $T_{\rm c2}$ ($T_{\rm c3}$) the two components $\eta_{+}$ and $\eta_{-}$ coexist, symbolically denoted by
A$_1$+A$_2$. 

Note that because  their transition temperatures are different,
A$_1$+A$_2$ is not the so-called A-phase which is unitary, but generically non-unitary
except at the degenerate point $M_a$=0 where the totally symmetric phase is realized with time reversal symmetry preserved.
Likewise below $T_{\rm c3}$ all the components coexist; A$_1$+A$_2$+A$_0$ realizes.

Under an applied field with the vector potential $\bf A$, the gradient GL energy is given 


\begin{eqnarray}
F_{grad}=\sum_{\nu=a,b,c}\{K_a|D_x\eta_{\nu}|^2+K_b|D_y\eta_{\nu}|^2+K_c|D_z\eta_{\nu}|^2\}
\label{gradient}
\end{eqnarray}

\noindent
where $D_j=-i\hbar \partial_j-2eA_j/c$ and the mass terms are characterized by the coefficients $K_j$ ($j=a,b,c$) in D$_{2h}$.
We emphasize here as seen from this form of Eq.~(\ref{gradient}) 
that $H_{\rm c2}$ for the three components each starting at $T_{\rm c j}$ ($j=1,2, 3$) intersects
each other, never avoiding or leading to a level repulsion. The level repulsion may occur 
for the pairing states belonging to multi-dimensional representations
 (see for example [\onlinecite{repulsion1,repulsion2,repulsion3,repulsion4}] in UPt$_3$).
The external field $H$ implicitly comes into $T_{\rm c j}$ ($j=1,2, 3$) through $M_a(H)$ 
in addition to the vector potential $A$.
This gives rise to the orbital depairing mentioned above.

The magnetic coupling $\kappa$, which is a key parameter to characterize materials of interest
in the following, is estimated~\cite{mermin} by

\begin{eqnarray}
\kappa=T_{\rm c}{N'(0)\over N(0)}ln(1.14\omega/T_{\rm c})
\label{kappa}
\end{eqnarray}

\noindent
where $N'(0)$ is the energy derivative of the normal density of states $N(0)$ at the Fermi level 
and $\omega$ is the energy cut-off.
This term arises from the electron-hole asymmetry near the Fermi level. $\kappa$ indicates
the degree of this asymmetry, which can be substantial for a narrow band.
Thus the Kondo coherent band in heavy Fermion
materials, such as in our case, is expected to be important.
We can estimate
$N'(0)/N(0)\sim 1/E_{\rm F}$ with the Fermi energy $E_{\rm F}$.
Because $T_{\rm c}$=2mK and $E_{\rm F}$=1K in superfluid $^3$He, $\kappa\sim10^{-3}$.
In the present compounds $T_{\rm c}\sim$1K and $E_{\rm F}\sim T_{\rm K}$ with the $T_{\rm K}$ 
Kondo temperature being typically~\cite{aokireview} 10$\sim$50K.
Thus $\kappa$ is much larger than that of superfluid $^3$He and is an order of $1\sim10^{-1}$.
We will assign the $\kappa$ value for each compound to reproduce the phase diagram in the following
as tabulated in Table I.

\subsection{Quasi-particle spectrum for general triplet state}

If we choose ${\vec \eta}'=\eta_b {\hat b}$ and ${\vec \eta}''=\eta_c {\hat c}$  with $\eta_a=0$ for the magnetic easy  $a$-axis, 
the quasi-particle spectra are calculated by

\begin{eqnarray}
E_{k,\sigma}=\sqrt{\epsilon(k)^2+(|{\vec \eta}|^2\pm|{\vec \eta}\times{\vec \eta}^{\star}|)\phi(k)^2}
\label{spectrum}
\end{eqnarray}

\noindent
or

\begin{eqnarray}
E_{k,\sigma}=\sqrt{\epsilon(k)^2+\Delta_{\sigma}(k)^2},
\label{spectrum2}
\end{eqnarray}

\noindent
where the gap functions for two branches are


\begin{eqnarray}
\Delta_{\uparrow}(k)&=&|\eta_b+\eta_c|\phi(k)\nonumber\\
\Delta_{\downarrow}(k)&=&|\eta_b-\eta_c|\phi(k)\nonumber\\
\Delta_{0}(k)&=&|\eta_a|\phi(k).
\label{spectrum3}
\end{eqnarray}

\noindent
Note that 
if $|\eta_c|=0$, $\Delta_{\uparrow}(k)=\Delta_{\downarrow}(k)$, which is nothing but the A-phase~\cite{leggett}.
When $|\eta_b|=|\eta_c|$, $\Delta_{\uparrow}(k)\neq0$ and $\Delta_{\downarrow}(k)=0$,
which is the non-unitary A$_1$ phase for $\eta_a=0$. 
The gap in one of the two branches vanishes and the other remains ungapped.
Therefore, if we assume that in the normal state $N_{\uparrow}(0)=N_{\downarrow}(0)$,
the A$_1$ phase is characterized by having the ungapped DOS $N_{\downarrow}(0)=N(0)/2$
with $N(0)=N_{\uparrow}(0)+N_{\downarrow}(0)$.
Generically, however, since $N_{\uparrow}(0)\neq N_{\downarrow}(0)$, that is, $N_{\uparrow}(0)> N_{\downarrow}(0)$
in the A$_1$ phase,  which is energetically advantageous than the A$_2$ phase, the ``residual DOS'' is equal to 
$N_{\downarrow}(0)$, which is likely less-than-half rather then more-than-half physically.
In the non-unitary state with the complex $\bf d$-vector,
the time reversal symmetry is broken.

In the most general case where all components $\eta_a$, $\eta_b$, and $\eta_c$ are no-vanishing,
the quasi-particle spectra are calculated by diagonalizing the $4\times4$ eigenvalue matrix.
Namely in terms of Eq.~(\ref{spectrum3}) the spectrum is given by

\begin{eqnarray}
E^2_{k}=\epsilon(k)^2+{1\over2}{\bigl\{}\Delta^2_{\uparrow}(k)+\Delta^2_{\downarrow}(k)+2\Delta^2_{0}(k)\nonumber\\
\pm\sqrt{(\Delta^2_{\uparrow}(k)-\Delta^2_{\downarrow}(k))^2+4\Delta^2_{0}(k)(\Delta^2_{\uparrow}(k)+\Delta^2_{\downarrow}(k))}{\bigr\}}.
\label{spectrum}
\end{eqnarray}

\noindent
It is easy to see that this spectrum is reduced to Eq.~(\ref{spectrum2}) when $\Delta_0(k)=0$.
This spectrum characterizes the phase $A_1+A_2+A_0$ realized in UTe$_2$ under pressure
as we will see shortly.

\section{Prototypes of phase transitions}

\begin{figure}
\includegraphics[width=8cm]{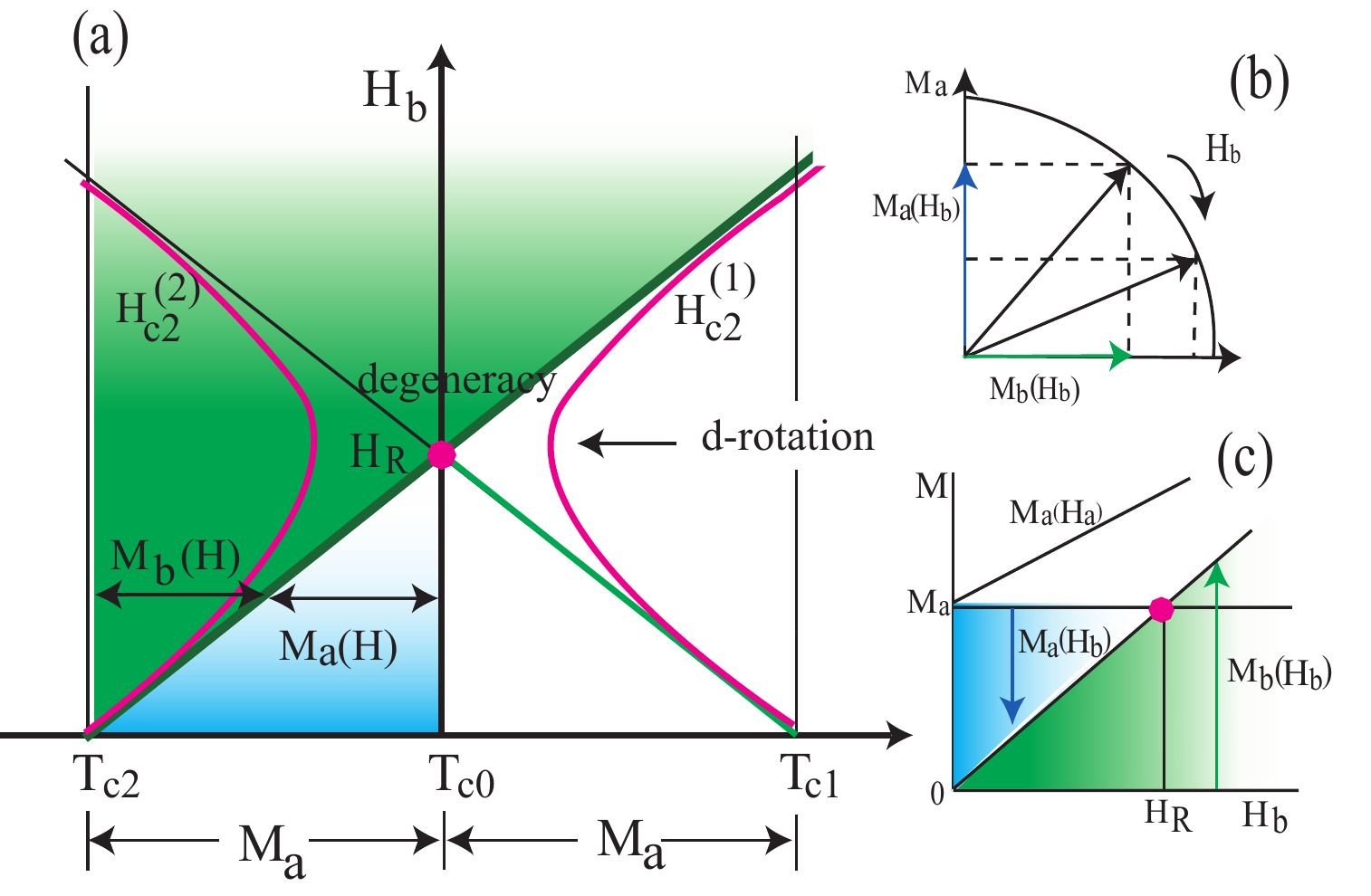}
\caption{(color online)
(a) Prototype phase diagram in the $T$ and $H_b$ plane
where $H_b$ is parallel to the magnetic hard $b$-axis and the moment $M_a$ points to the
easy  $a$-axis. The two transition lines of $T_{c1}$ and $T_{c2}$ (red curves) initially decreases and increases
respectively as $H_b$ increases toward the degeneracy point at $T_R$. There the projection of the FM moment $M_a$ vanishes.
During this process, by rotating the $\bf d$-vector to catch the magnetization $M_b(H_b)$ (the green lines) instead of $M_a(H_b)$,
$H^{(1)}_{\rm c2}$ and $H^{(2)}_{\rm c2}$ turn around their directions.
(b) Under the perpendicular field $H_b$ the spontaneous moment $M_a$ rotates toward the $b$-direction.
The projection $M_b(H_b)$ of $M_a$ on the $b$-axis increases.
(c) The rotation field $H_R$ is indicated as the red dot where $M_b(H_b)=M_a(H=0)$.
}
\label{proto}
\end{figure}

Let us now consider the action of the external field $H_b$ applied to the magnetic hard $b$-axis on the FM moment $M_a$, 
pointing parallel to the $a$-axis.
The $a$-axis component of the moment $M_a(H_b)$ generally decreases 
as it rotates toward the $b$-axis as shown in Fig.~\ref{proto}(b).
As discussed in the next section in more details based on experimental data, 
it is observed in URhGe through the neutron experiment~\cite{levy}. 
Here we display the generic and typical magnetization curves of $M_a$
and $M_b$ in Fig.~\ref{proto}(c) where $H_R$ denotes a characteristic field for $M_b(H_b)=M_a(H_b=0)$.
The induced moment $M_b$ reaches the spontaneous FM moment $M_a$ at zero field by rotating the FM moment, implying that
the FM moment points to the $b$-axis above $H_R$.
Experimentally, it is realized by the so-called meta-magnetic transition via a first order transition in URhGe~\cite{levy}
 and UTe$_2$~\cite{miyake} or 
gradual change in UCoGe~\cite{knafoCo}.

As displayed in Fig.~\ref{proto}(a), by increasing 
$H_b$, $T_{c1}$ ($T_{c2}$)  decreases (increases) according to Eq.~(\ref{tc}).
The two transition lines $T_{c1}(H_b)$ =$T_{c2}(H_b)$
meet at $H_b=H_R$. As $H_b$ is further increased, $T_{c1}$ also increases
by rotating the $\bf d$-vector direction such that the $\bf d$-vector becomes perpendicular to ${\bf M}_b$, 
which maximally gains the magnetic coupling energy
$i\kappa {\vec M}\cdot {\vec \eta}\times {\vec \eta}^{\star}$ in Eq.~(\ref{f2}).
This process occurs gradually or suddenly, depending on the situations of the magnetic subsystem
and the spin-orbit coupling that locks the $\bf d$-vector to the underlying lattices.
Therefore $H_R$ may indicate simultaneously the $\bf d$-vector rotation.
It should be noted, however, that if the spin-orbit coupling is strong, the $\bf d$-vector rotation is prevented.
In this case $H^b_{\rm c2}$ exhibits a Pauli limited behavior as observed in UTe$_2$ under pressure~\cite{aokiP}.

\begin{figure}
\includegraphics[width=8cm]{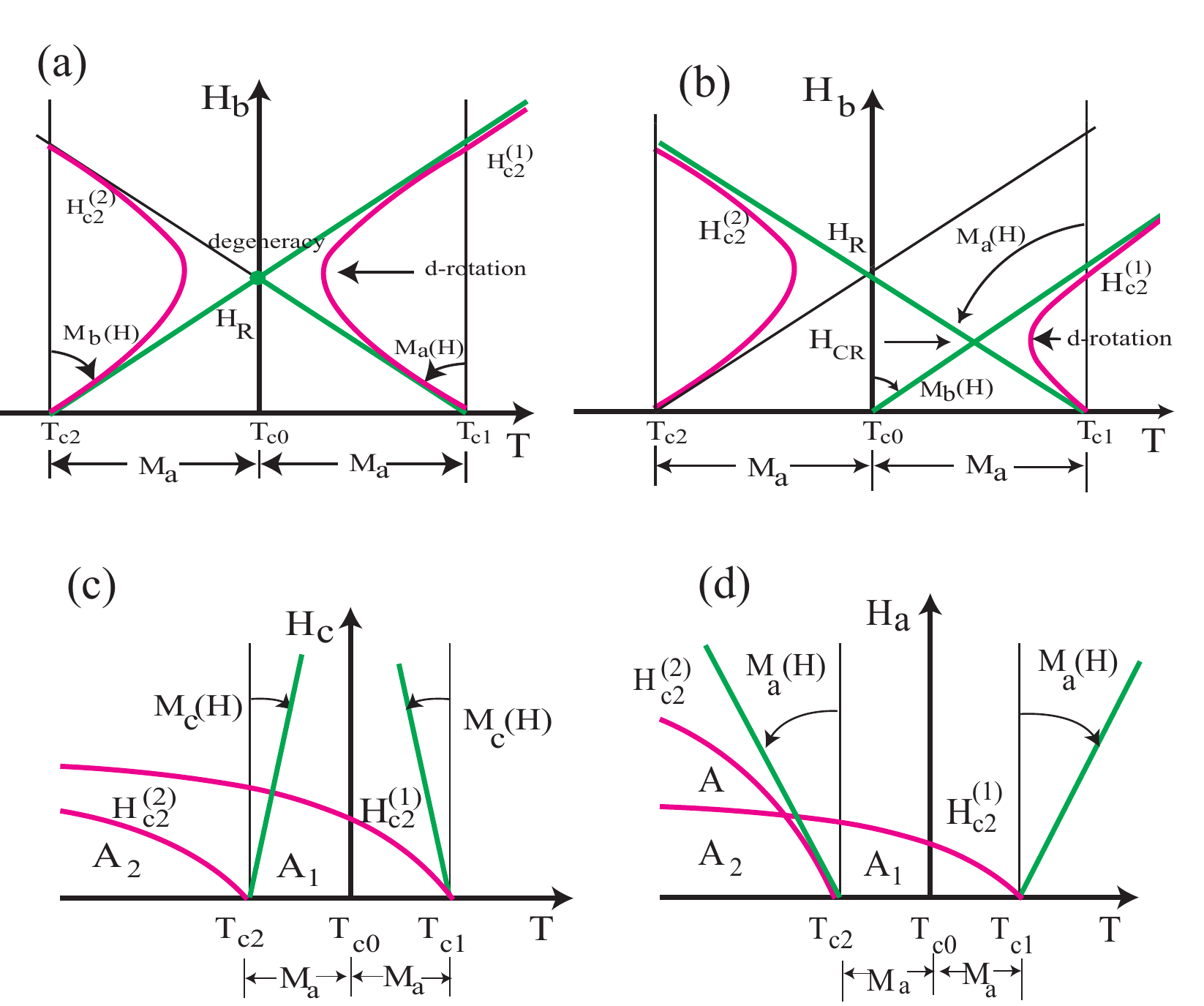}
\caption{(color online) 
Two types (a) and (b) of the phase diagram for
 $H\parallel b$ with the $b$-axis (hard axis).
(a) is the same as in Fig.~\ref{proto}(a).
(b) At $H_{\rm CR}$ defined by $M_b(H_b)=M_a(H_b)$, $H^{(1)}_{\rm c2}$ turns round by rotating the $\bf d$-vector to catch $M_b$
starting from $T_{c0}$.
(c) $H\parallel c$ with the $c$-axis (another hard axis).
(d) $H\parallel a$ with the $a$-axis (easy axis). 
The green lines are the respective 
magnetization curves and
the red curves are $H^{(1)}_{\rm c2}$ and $H^{(2)}_{\rm c2}$.
}
\label{prototype123}
\end{figure}

In Fig.~\ref{prototype123} we show prototype phase diagrams for different situations.
In addition to that displayed in Fig.~\ref{prototype123}(a), which is the same as in Fig.~\ref{proto}(a),
there is the case in which $T_{c1}$ is bent before reaching $H_R$ as shown in Fig.~\ref{prototype123}(b).
The magnetization curve $M_b(H_b)$ starting at $T_{c0}$ exceeds the decreasing $M_a$
at a lower field $H_{\rm CR}$ defined by $M_b(H_b)=M_a(H_b)$. $H^{(1)}_{\rm c2}$ turns around there by rotating the $\bf d$-vector.
We will see this case in the following analysis.

In the $H_c$ case for the field direction parallel to the another hard $c$-axis the phase diagram is shown in Fig.~\ref{prototype123}(c).
Since $H_c$ does not much influence on $M_a(H_c)$, both $H^{(1)}_{\rm c2}$ and $H^{(2)}_{\rm c2}$ are suppressed
by the orbital depression of $H_c$. 
When magnetic field is applied to the magnetic easy $a$-axis, the spontaneous moment $M_a(H_a)$
increases monotonically, as shown in Fig.~\ref{prototype123}(d).  According to Eq.~(\ref{tc}), $T_{c1}$ ($T_{c2}$)
increases (decreases) as $H_a$ increases. Thus, theoretically $H_{c2}^{(1)}$ can have a positive slope at $T_{c1}$. 
However, the existing data on UCoGe~\cite{wu} indicate that it is negative as seen shortly.
This is because the strong 
orbital depairing $H'^0_{\rm c2}$ overcomes the positive rise of $T_{c1}$.
 Moreover, $H_{c2}^{(2)}$ is strongly suppressed by both $T_{c2}$ and the orbital
 effect $H'^0_{\rm c2}$, resulting in a low $H^a_{c2}$, compared with $H^b_{c2}$.
 This $H_{c2}$ anisotropy is common in these compounds~\cite{aokireview}.
 From the above considerations, the enhanced $H^b_{c2}$ is observed because the
 higher part of the field in $H_{c2}$ belongs to $H_{c2}^{(2)}$, which has a positive slope.

Within the GL scheme it is easy to estimate $H_{\rm c2}$ as follows.
We start with the $H_{\rm c2}$ expression:

\begin{equation}
H_{\rm c2}(T)=A_0\{T_c(H_{\rm c2})-T\}
\label{Hc2}
\end{equation}

\noindent
with $A_0={\Phi_0\over 2\pi\hbar^2}4m\alpha_0$, $m$ effective mass, and $\Phi_0$ quantum unit flux.
 Here $T_c$ depends on $H$ though $M_a(H)$ as described above.
Thus, the initial slope of $H'_{\rm c2}$ at $T_c$ is simply given by

\begin{equation}
H'_{\rm c2}(T)=A_0{dT_c\over dH_{\rm c2}}H'_{\rm c2}-A_0.
\label{Hc2prime}
\end{equation}

\noindent
It is seen that if ${dT_c/dH}=0$ for the ordinary superconductors, 
$H'^0_{\rm c2}(T)=-A_0<0$. The slope 
$H'_{\rm c2}(T)$ is always negative.
However, Eq.~(\ref{Hc2prime}) is expressed as


\begin{equation}
H'_{\rm c2}(T)={-A_0\over {1-A_0({dT_c\over dH_{\rm c2}}})},
\end{equation}

\noindent
or




\begin{eqnarray}
{1\over |H'_{\rm c2}|} &=& {1\over |H'^0_{\rm c2}|}+|{dT_{\rm c} \over dH_{\rm c2}}| \nonumber \\
&=& {1\over |H'^0_{\rm c2}|}+{1\over |{dH_{\rm c2}\over dT_{\rm c}(H)}|}.
\label{parallel}
\end{eqnarray}

\noindent
The condition for attaining the positive slope, $H'_{\rm c2}(T)>0$ implies 
$|H'^0_{\rm c2}|>({dH\over dT_{\rm c}})$ at $H_{\rm c2}$.
This is a necessary condition to achieve S-shaped or L-shaped $H_{\rm c2}$ curves experimentally observed.
This is fulfilled when $|H'^0_{\rm c2}|$ is large enough, that is, the orbital depairing is small, 
$|{dT_c\over{dH}}|$ at $H_{\rm c2}$ is large, or the $T_c$ rise is strong enough.

It is noted that when $1-A_0({dT_c\over dH_{\rm c2}})=0$, the $H_{\rm c2}(T)$ curve has a 
divergent part in its slope, which is observed in UCoGe as a part of the S-shape. 
It is clear from the above that when $dT_c/dH<0$,
$|H'_{\rm c2}(T)|<|H'^0_{\rm c2}|$ because  the two terms in Eq.~(\ref{parallel}) are added up to further depress $H_{\rm c2}(T)$.
In this case the slope $|H'_{\rm c2}|$ is always smaller than the original $|H'^0_{\rm c2}|$
as expected.

\begin{figure}
\includegraphics[width=8.5cm]{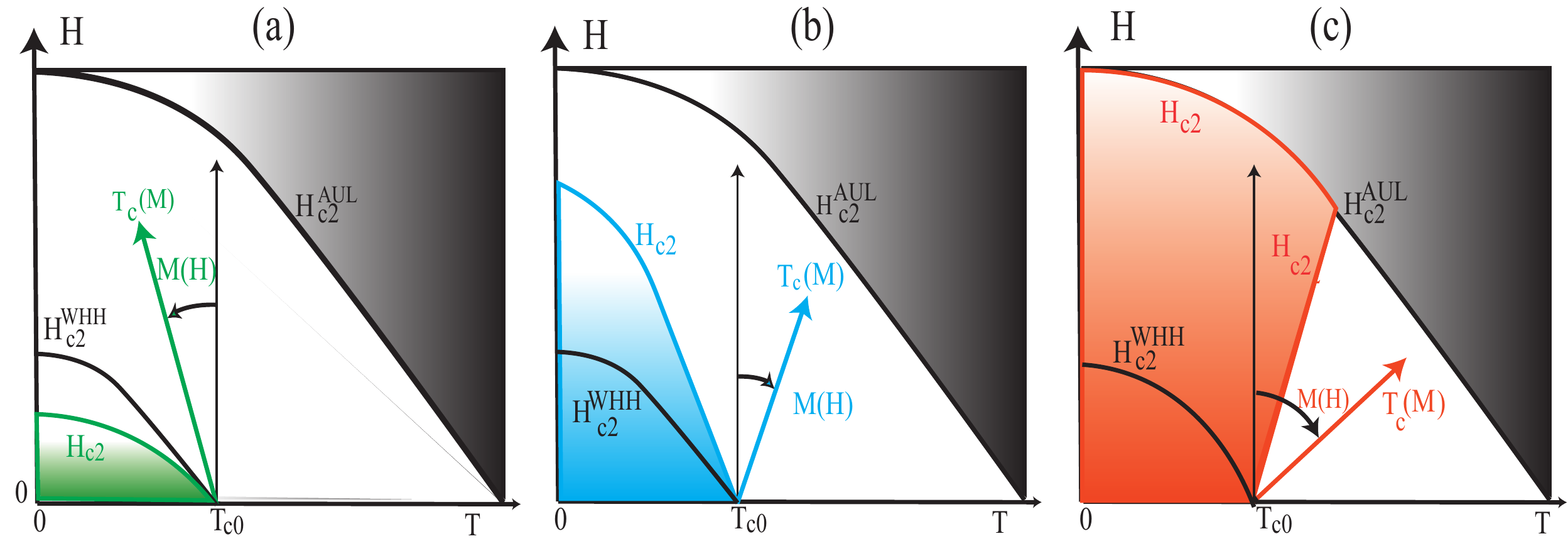}
\caption{(color online) 
$H_{\rm c2}$ changes due to the competition between the orbital depairing and $T_c(M)$.
(a) $T_c(M)$ decreases as a function of the applied field $H$. The orbital depairing is added up to  further depress 
$H_{\rm c2}$ than $H^{\rm WHH}_{\rm c2}$. 
(b) $T_c(M)$ increases as a function of the applied field $H$, competing with the orbital depairing. 
The resulting $H_{\rm c2}$ is enhanced compared with $H^{\rm WHH}_{\rm c2}$.
(c) $T_c(M)$ increases strongly as a function of the applied field $H$.
$H_{\rm c2}$ has a positive initial slope and keeps growing until hitting the absolute upper limit $H^{\rm AUL}_{\rm c2}$.
Then $H_{\rm c2}$ follows this boundary.
}
\label{Hc2graph}
\end{figure}

In Fig.~\ref{Hc2graph} we show the changes of $H_{\rm c2}$ when the competition between the 
orbital suppression and $T_c(M)$ varies. We start from the orbital limited $H^{\rm WHH}_{\rm c2}$
curve with $T_c$ unchanged as a standard one.
When $T_c(M)$ decreases with increasing $H$, the resulting $H_{\rm c2}$
is further suppressed compared with $H^{\rm WHH}_{\rm c2}$  as shown in Fig.~\ref{Hc2graph}(a).
$T_c(M)$ as a function of $H$ through $M(H)$ becomes increasing
as shown in Fig.~\ref{Hc2graph}(b), $H_{\rm c2}$ is enhanced compared to $H^{\rm WHH}_{\rm c2}$,
exceeding the $H^{\rm WHH}_{\rm c2}$ value.
Figure.~\ref{Hc2graph}(c) displays the case where $T_c(M)$ increases stronger than that in Fig.~\ref{Hc2graph}(b),
$H_{\rm c2}$ has a positive slope and keeps increasing until it hits the upper limit $H^{\rm AUL}_{\rm c2}$. 
There exists the absolute upper limit (AUL) for $H_{\rm c2}$.
Even though $T_c(M)$ keeps increasing with increasing $M(H)$, $H_{\rm c2}$
terminates at a certain field because a material has its own coherent length $\xi$
which absolutely limits $H^{\rm AUL}_{\rm c2}=\Phi_0/2\pi\xi^2$. Beyond  $H^{\rm AUL}_{\rm c2}$
there exists no superconducting state.


There could be several types of $H^b_{\rm c2}$ curves for $H$ applied to the $b$-axis (hard axis),
depending on several factors: 

\noindent
(A) the magnitude of the spontaneous 
moment $M_a$, 

\noindent
(B) its growth rate against $H_b$, 

\noindent
(C) the coupling constant $\kappa$, 

\noindent
(D) the relative position of $T_{c0}$ on the temperature axis.

\begin{figure}
\includegraphics[width=9cm]{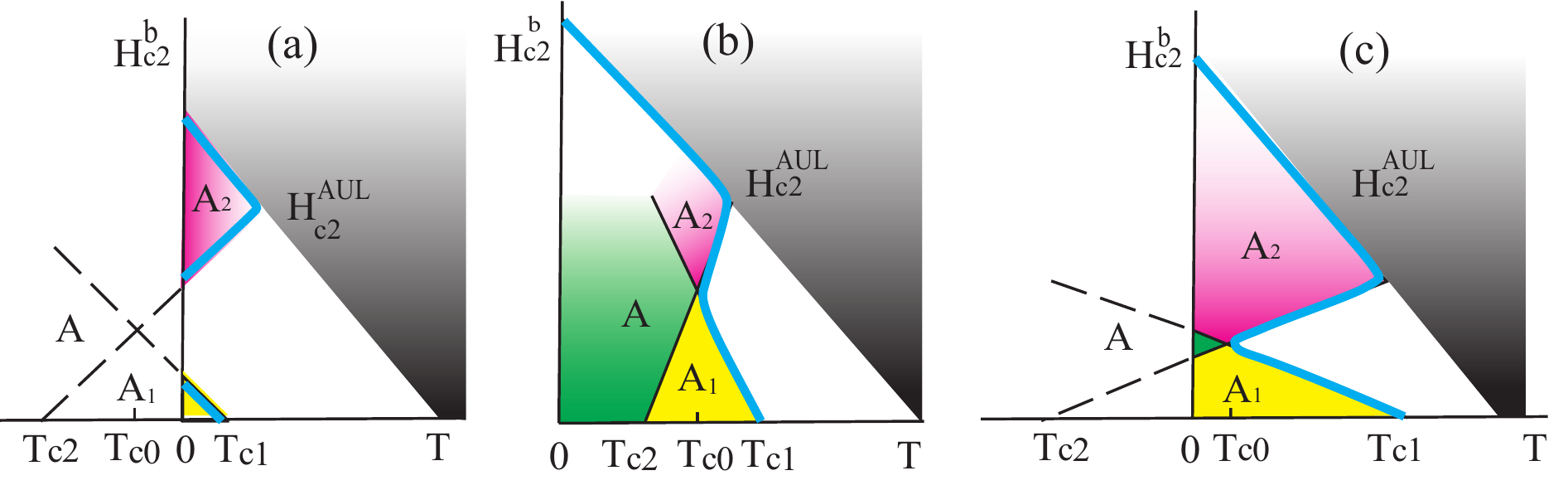}
\caption{(color online) 
Schematic typical phase diagrams for $H$ parallel to the $b$-axis with $A_1$, $A_2$ and $A$ phases,
whose structure depends on the position of $T_{c0}$ and the separation of $T_{c1}$ and $T_{c2}$.
The absolute upper limit $H_{\rm c2}^{\rm AUL}$ is indicated as the grey region.
(a) The reentrant SC situated at high fields such as in URhGe.
(b) S-shape $H_{\rm c2}$ with the double transitions from the $A_1$ to the $A$ phase such as in UCoGe.
(c) L-shape $H_{\rm c2}$ where the high field phase is the $A_2$ phase such as in UTe$_2$.
}
\label{protoHc2b}
\end{figure}

\noindent
Possible representative $H^b_{\rm c2}$ curves are displayed in Figs.~\ref{protoHc2b}(a), ~\ref{protoHc2b}(b) and~\ref{protoHc2b}(c).

When the hypothetical $T_{c0}$ is situated in the negative temperature side, the realized phase is only the A$_1$-phase at $H_b=0$.
In high field regions SC reappears as the reentrant SC (RSC) by increasing $M_b(H_b)$, which is shown in Fig.~\ref{protoHc2b}(a).
The reentrant SC is separated from the lower SC.

As shown in Fig.~\ref{protoHc2b}(b) the two transition temperatures, $T_{c1}$  and $T_{c2}$ are realized at $H_b=0$,
that is,  it shows double transitions at zero field,  giving rise to the $A_1$ and $A_2$ phases.
The three phases $A_1$, $A_2$ and $A$ appear in a finite $H_b$ region. $H_{\rm c2}$ could have an S-shape.
This corresponds to either Figs.~\ref{prototype123}(a) or (b).

When the separation between $T_{c1}$  and $T_{c2}$ becomes wider because of  increasing the spontaneous moment $M_a$
and/or the larger magnetic coupling $\kappa$,  $H_{\rm c2}$ has an L-shape as displayed in Fig.~\ref{protoHc2b}(c).
This could happen also when the moment rotation field $T_R$ is situated at relatively lower field than the overall 
$H_{\rm c2}$.

In the following we discuss those typical $H_{\rm c2}$ behaviors based on the realistic magnetization curves for
each compound, reproduce the observed $H_{\rm c2}$ curves and predict the existence of the multiple phase diagram.

\section{Magnetization curves}

In order to understand their peculiar $H_{\rm c2}$ shapes 
and resulting pairing symmetry in three compounds, it is essential to
know their magnetic responses to applied magnetic fields as mentioned above.
Here we analyze their magnetism and estimate the magnetization curves of
the spontaneous moment under the transverse field, which is not probed
by conventional magnetization measurements. In the following 
we consider the cases of URhGe and UCoGe, and UTe$_2$ with the $c$-axis and $a$-axis are the easy axes respectively
as tabulated in Table I.
We mainly discuss URhGe as an typical example.
The concepts introduced here are applied to the other systems with appropriately changing the
notation for the magnetic easy axis.

 \begin{table}[t]
  \caption{Magnetic properties and $\kappa$ values}
  \label{table:data_type}
  \centering
  \begin{tabular}{ccccc}
    \hline
    materials  & Curie temp.[K] & easy axis &  moment[$\mu_B$] & $\kappa$[K/$\mu_B$]\\
    \hline \hline
    URhGe  &9.5 & $c$-axis & M$_c$=0.4  &  2.0 \\
    UCoGe & 2.5 & $c$-axis  & M$_c$=0.06 &1.8\\
     UTe$_2$ & --  & $a$-axis & $\sqrt{\langle{\rm M}_a^2\rangle}$=0.48 & 6.9\\
           \hline
  \end{tabular}
\end{table}

\subsection{Rigid rotation picture: Spontaneous moment rotation}

\begin{figure}
\includegraphics[width=4cm]{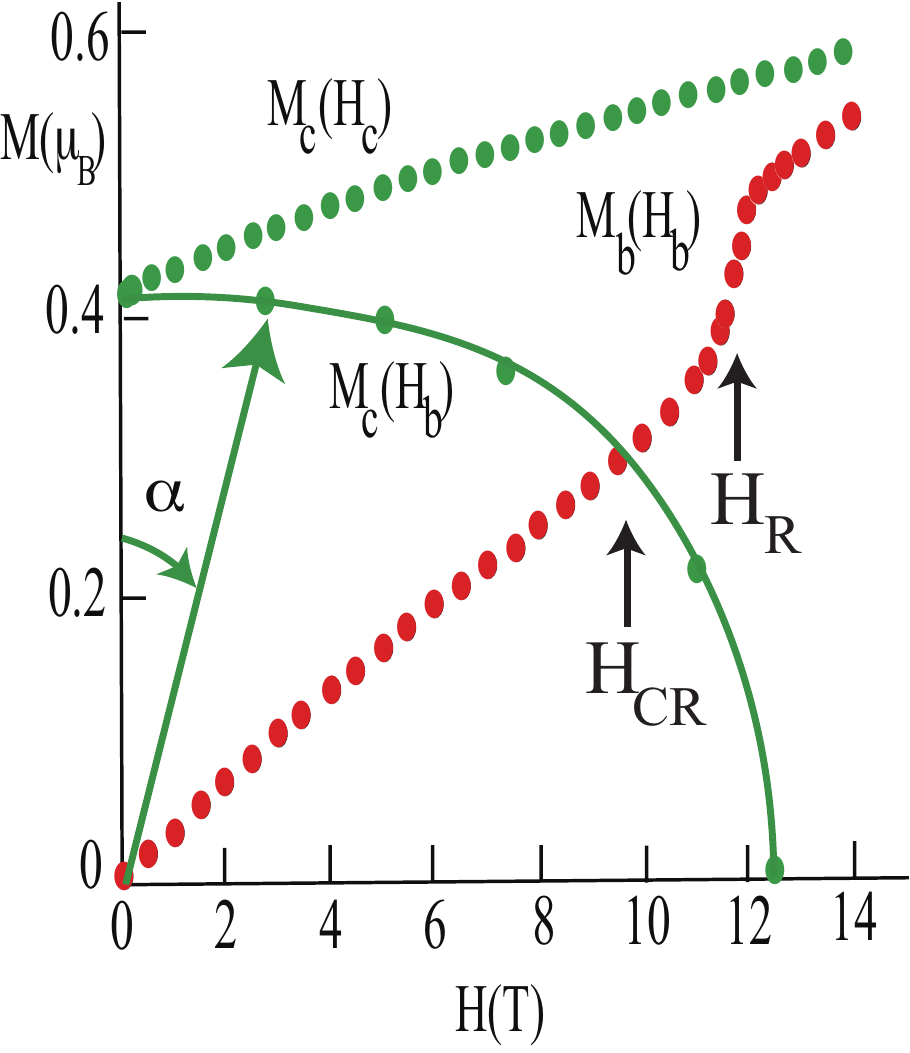}
\caption{(color online) 
The ferromagnetic spontaneous moment  $M_c(H_b)$ rotation indicated by the green arrow under the field $H\parallel b$ in URhGe.
At $H_b=H_R$, it completely orients along the $b$-axis direction via a first order transition
where $M_b(H_b)$ shows a jump of the magnetization. $H_{CR}$ is defined by the field $M_c(H_b)=M_b(H_b)$.
The rotation angle $\alpha$ from the $c$-axis is measured by
neutron experiment [\onlinecite{levy}]. The magnetization curves $M_b(H_b)$ and $M_c(H_c)$ are from [\onlinecite{hardy}].
}
\label{Mc}
\end{figure}

When the applied field $H_b$ is directed to the hard axis, or the $b$-axis, the spontaneous moment
$M_c(H_b)$ pointing to the $c$-axis in URhGe  rotates gradually toward  the applied field direction.
At around $H_R=12$T, $M_c(H_b)$ quickly turns to the $b$-direction by rotating the moment as shown in Fig.~\ref{Mc}.
We define the crossing field $H_{\rm \rm CR}$ at which $M_c(H_b)=M_b(H_b)$.
Note that $H_R$ and $H_{\rm \rm CR}$ are different concepts as  is 
clear from Fig.~\ref{Mc}  and also in UCoGe where $H_{\rm \rm CR}\sim $ a few T and $H_R$=45T~\cite{knafoCo}.
Simultaneously and correspondingly, the $M_b(H_b)$ moment jumps via a first order transition.
Above $H_b>H_R$ the spontaneous moment is completely aligned along the $b$-axis as seen from Fig.~\ref{Mc}.
This phenomenon is often called as the meta-magnetic transition. But this is just the moment rotation
since it is demonstrated that the total magnetization $\sqrt{M^2_c(H_b)+M_b^2(H_b)}$ hardly changes 
and remains a constant during this first order transition process~\cite{levy}. 

This implies that $M_c(H_b)=M_c\cos(\alpha(H_b))$, and $M_b(H_b)=M_c\sin(\alpha(H_b))$ with $\alpha(H_b)$ being the
rotation angle of $M_c(H_b)$ from the $c$-axis.
The rotation angle $\alpha(H_b)$ is accurately measured by the neutron scattering experiment by L\'{e}vy, et al\cite{levy}
who construct the detailed map of the rotation angle in the $H_b$ and $H_c$ plane.
This rotation process is mirrored by the magnetization curve of $M_b(H_b)$ so that
the projection of $M_c(H_b)$ onto the $b$-axis manifests itself on $M_b(H_b)$ as shown in Fig.~\ref{Mc}.
The crossing of $M_b(H_b)$ and $M_c(H_b)$ occurs around at $H_{\rm CR}=9\sim10$T,
corresponding to roughly $M_c(H_b)/\sqrt2 \sim M_b(H_b)$. That is, $M_c(H_b)$ rotates by
the angle $\alpha=45^{\circ}$ from the $c$-axis at $H_{\rm \rm CR}$.
This first order phase transition phenomenon in URhGe under the transverse field is neatly described by Mineev~\cite{mineev}
using the GL theory. This is within more general framework of the so-called meta-magnetic transition theory
based on the GL phenomenology~\cite{wohlfarth,shimizu,yamada} for itinerant ferromagnets.

Those considerations based on the experimental facts demonstrate to
hold ``a rigid moment rotation picture''.
We assume this picture applicable to the other compounds too.

\subsection{Extraction of the $M_b$ moment for the tilted fields from the $b$-axis data}

\begin{figure}
\includegraphics[width=8cm]{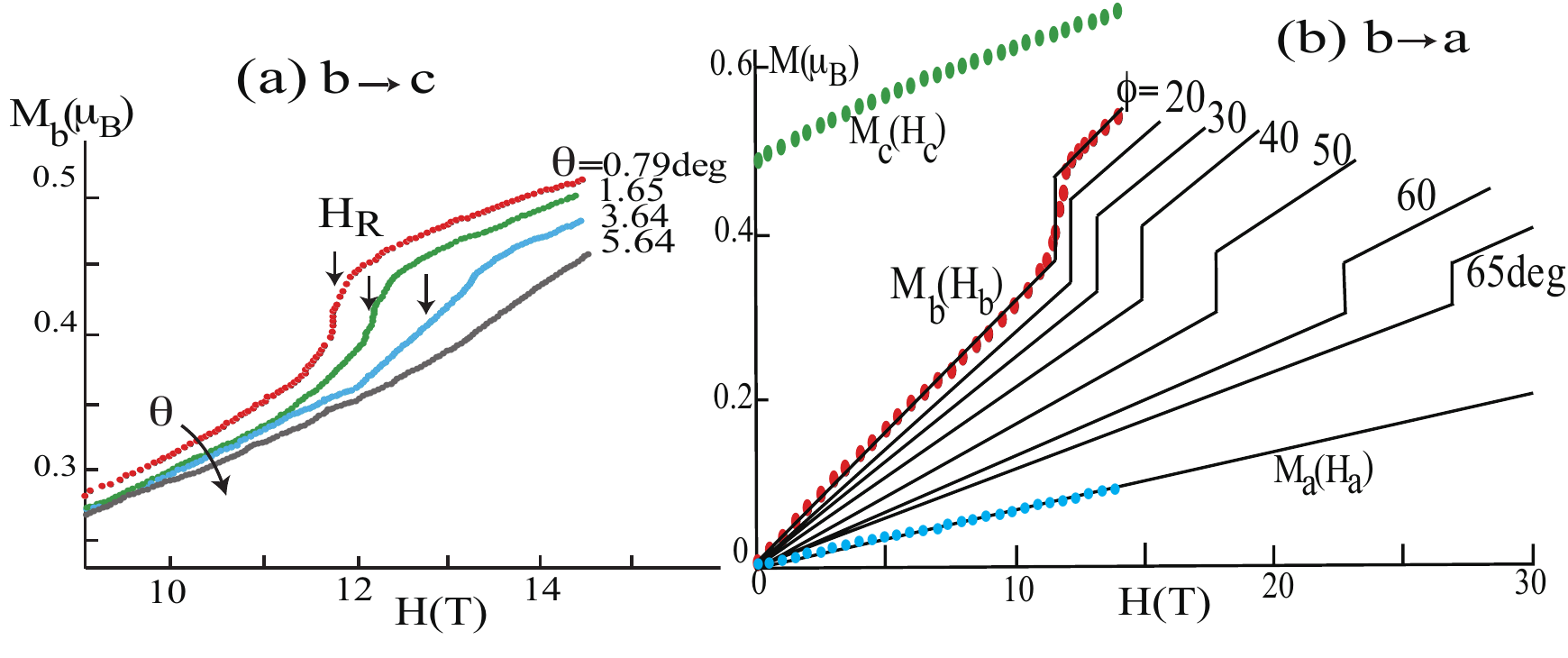}
\caption{(color online) 
(a) The magnetization component of $M_b(H)$ in URhGe under the field direction tilted from the $b$-axis toward the $c$-axis by $\theta$,
estimated from the experimental data of $M(H)$ [\onlinecite{nakamura}].
(b) The magnetization $M(H)$ in URhGe under the field direction tilted from the $b$-axis toward the $a$-axis by $\phi$
estimated from the experimental data (dots) of $M(H)$ [\onlinecite{hardy}],
including magnetization curves for three $a$, $b$ and $c$-directions for reference.
}
\label{thetaphi}
\end{figure}

When the applied field direction is rotated from the $b$-axis toward the easy axis $c$ by 
the angle $\theta$, the magnetization curves are measured by Nakamura, et al~\cite{nakamura}.
It is obvious that the measured magnetization $M(\theta)$ contains the contribution from the 
spontaneous moment ${\bf M}_c$ projected onto the applied field direction, that is, $M_c\sin(\theta)$.
This is confirmed experimentally at least lower fields up to  $H<5$T and $T$=2K~\cite{aokiprivate}.
Thus in this situation, we can extract the $M_b(H)$ curves by simply subtracting the contribution 
$M_c\sin(\theta)$ from the measured data~\cite{nakamura}. The result is shown in Fig.~\ref{thetaphi}(a).
It is seen that by increasing the angle $\theta$, the first order transition field $H_R$ shifts to higher fields
and the jump gets smaller compared to the $b$-axis case, reflecting that the moment projection onto the applied field
direction decreases. This method is valid only for the small angle $\theta$ and relatively small field regions
because here the $M_c$ moment is assumed to be fixed under the action of small field component along the
$c$-axis.

It may be difficult to extract reliably the $M_b(H)$ information for further high fields
even though the tilting angle is small, and also for larger angles $\theta$.
There are two factors to be taken into account, which are internally related:
One is that the $c$-component magnetic field acts to prevent the moment from further rotating it
toward the $b$-axis upon increasing tilting field $H$ by $\theta$ from the $b$-axis. 
This ``rotation angle locking effect'' becomes important
for the field just before $H_R(\theta)$ where the moment ultimately rotates completely along the $b$-axis
in the higher fields.
The other factor to be considered is the modification of the free energy landscape of the $M_b$ versus $M_c$ space.

As mentioned, the first order transition of the moment rotation is described by Mineev~\cite{mineev} who considers the
competition between the ferromagnetic state at $M_c$ and the paramagnetic state with $M_b$
stabilized by the Zeeman effect due to the external field $H_b$ within a GL free energy theory.
The transverse field $H_b$ necessarily destabilizes the second order FM phase transition at $H_R$ because $H_b$ contributes
negatively to the quartic term coefficient of $M_c^4$, giving rise to a first order transition.
The extra term coming from the tilting field helps to stabilize the ferromagnetic state,
preventing the first order transition, thus making $H_R$ to higher field and the magnetization
jump smaller. Thus it is not easy to extract reliably the $M_b(H)$ under this free energy
landscape modification.
In the followings, we confine our arguments for small $\theta$ and use approximate $M_b(H)$ forms, which are enough for our purposes
to understand the peculiar $H_{c2}$. 

\begin{figure}
\includegraphics[width=8cm]{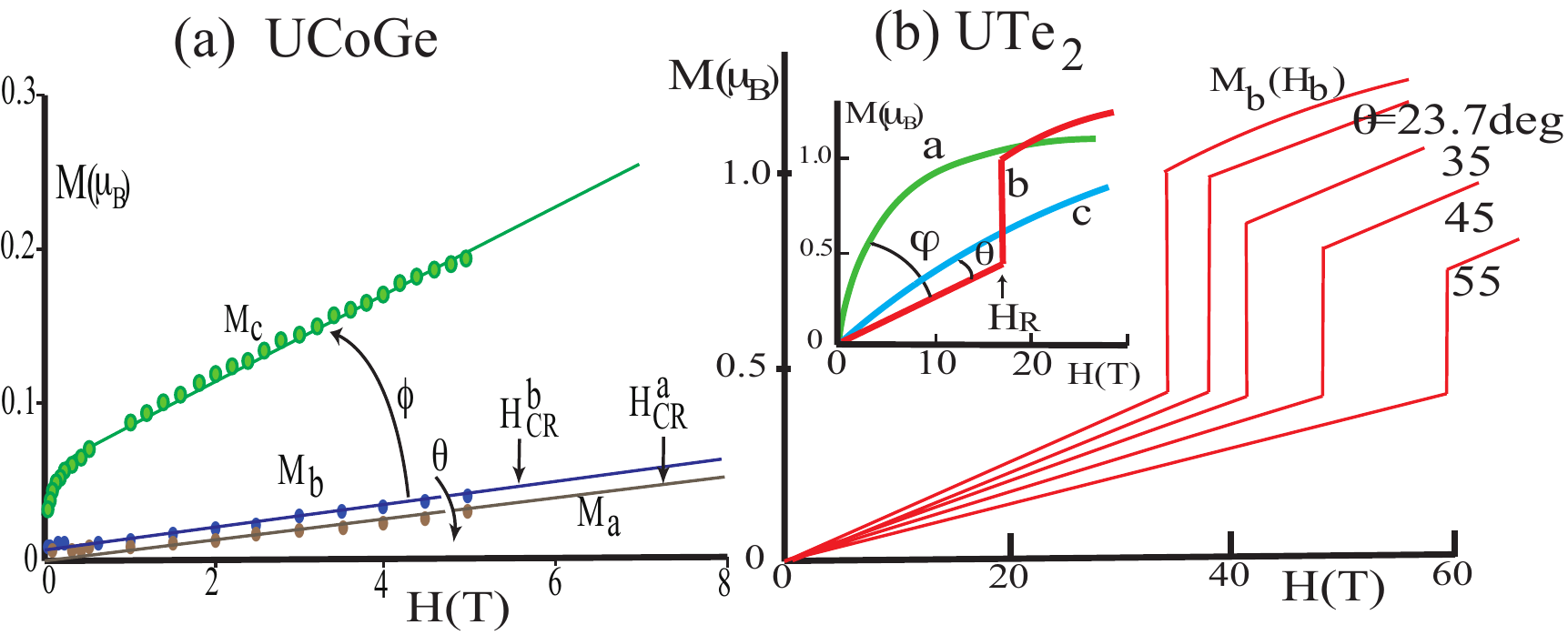}
\caption{(color online) 
(a) The magnetization curves for three $a$, $b$, and $c$-axes in UCoGe.
Here the crossing points $H^b_{\rm \rm CR}$ and $H^a_{\rm \rm CR}$ at which each curve surpasses the
spontaneous moment $M_c(H=0)=0.06\mu_B$.
(b) The magnetization curves of $M_b(H)$ for the field directions tilted from the $b$-axis toward the $c$-axis
by the angle $\theta$(degrees) in UTe$_2$. $\theta=23.7^{\circ}$ corresponds to $H\parallel(011)$ direction measured 
by [\onlinecite{miyakeprivate}].
Those are estimated by the method explained in the main text.
The inset shows the magnetization curves for three $a$, $b$, and $c$-axes in UTe$_2$.
$H_R$ is the first order transition for the moment rotation from the $a$-axis to the $b$-axis.
}
\label{Mall}
\end{figure}

\subsection{Applied field rotation from the $b$-axis to the hard axis}

In the case for the tilting angle $\phi$ from the $b$-axis toward the other hard axis $a$ of URhGe,
it is known~\cite{levy2} that $H_R(\phi)$ is scaled to $H_R(\phi)\propto 1/\cos(\phi)$, 
which is also the case in UTe$_2$~\cite{ran2}. This means that only the $M_b$ projection onto the 
$a$-axis matters to understand the magnetization process. Therefore, we can easily reconstruct the $M(\phi)$ by using the experimental data 
of $M_b(H_b)$ except for the fact that the induced $M_a(\phi)$ also contributes to $M(\phi)$.
This can be accomplished by an  ``elliptic formula'' derived as follows:

We start with $M_b(H_b)$ and $M_a(H_a)$ measured by usual magnetization experiments shown in Fig.~\ref{thetaphi}(b).
Assuming the linearity assumption: $M_b(\phi)=\chi_b H\cos(\phi)$ and $M_a(\phi)=\chi_a H\sin(\phi)$ with $\chi_i$ $(i=a, b)$ being
the magnetic susceptibility,
we add up the two components,



\begin{eqnarray}
M(\phi)&=&M_b\cos\phi+M_a\sin\phi \nonumber\\
&=&(\chi_b\cos^2(\phi)+\chi_a\sin^2(\phi))H \nonumber\\
&=&M_b(H_b)\cos^2\phi+M_a(H_a)\sin^2\phi.
\end{eqnarray}

\noindent
We call it an  ``elliptic formula''. 
Since the rotation field is given by

\begin{equation}
H_R(\phi)={H_R^b\over \cos(\phi)}
\end{equation}

\noindent
with $H_R^b$ the rotation field for the $b$-axis,
we obtain at $H=H_R$

\begin{equation}
M(\phi)=M_b(H_R){\bigl(}\cos\phi+{\chi_a\over \chi_b}\cdot{\sin^2\phi\over \cos^2\phi}\bigr).
\label{elliptic}
\end{equation}

\noindent
This formula gives the magnetization curve consisting of a straight line from $H=0$ up to $H_R$.
The magnetization jump at $H_R$ is calculated by projecting the jump $\delta M_b$ in $M_b(H_b)$, namely
$\delta M_b\cos(\phi)$.

The resulting reconstructions of $M(\phi)$ for various tilting angles are shown in Fig.~\ref{thetaphi}(b). 
By construction, when $\phi\rightarrow90^{\circ}$, $M(\phi)\rightarrow M_a(H_a)$.
We notice that the resulting $M(\phi)$ includes the contribution from $M_a$.
Those results should be checked experimentally
and will be used to reproduce the RSC in URhGe.
As shown in Fig.~\ref{Mall}(b) this idea is also applied to UTe$_2$ where the RSC appears centered around $\theta=$35$^{\circ}$
from the $b$-axis toward the another hard axis $c$.

As a final comment on the magnetization of UCoGe shown in Fig.~\ref{Mall}(a),
it should be mentioned that since $H_R\sim45$T~\cite{knafoCo}, for the following discussions on this system
the characteristic magnetic fields $H^b_{\rm \rm CR}\sim6$T and $H^a_{\rm \rm CR}\sim7$T are relevant to notice from this figure.
We also note that two magnetization curves $M_b$ and $M_a$ behave similarly.
It is anticipated that $H_{\rm c2}$ for the two directions should be resemble.
This is indeed the case as will be seen next.

\section{Application to experiments on three compounds}

Let us now examine the present theory to understand a variety of experiments 
on the three compounds, URhGe, UCoGe and UTe$_2$.
In order to clarify the essential points of the problem and for the discussions followed to be transparent,
and to minimize the free adjustable parameters, we take a simplified minimal version of the present theory.
It is quite easy to finely tune our theory by introducing
additional parameters such as $\beta_1$ and $\beta_2$ in the GL theory Eq.~(\ref{f2}) for each compound if necessary. 
We assume that

\begin{eqnarray}
T_{\rm c1}=T_{\rm c0} +{\kappa}M_a, \nonumber \\
T_{\rm c2}=T_{\rm c0} -{\kappa}M_a, \nonumber \\
T_{\rm c3}=T_{\rm c0} -bM^2_a
\label{tc3}
\end{eqnarray}

\noindent
for the spontaneous FM moment $M_a$ with the easy $a$-axis.
We have redefined $\kappa/\alpha_0$ as $\kappa$ and $b/\alpha_0$ as $b$, ignoring the correction 
in Eq.~(\ref{tc2}) from the higher order GL terms. 
Since $\kappa$ is a converter of the units from $\mu_B$ to K, we further simplify the notation in that
$\kappa M$ having the dimension of  temperature in [K] is denoted as $M$ in [K] in the following phase diagrams as mentioned before.
We use the $\kappa$ values for three compounds throughout this paper as shown in Table I
where the magnetic properties are also summarized.

In the following, we intend to produce the observed $H_{\rm c2}$ curves only qualitatively, not quantitatively.
This is because the experimental $H_{\rm c2}$ shapes somewhat  depend on the experimental methods.
For example, see Fig.~1  in Ref.~[\onlinecite{wu1}] where $H_{\rm c2}$ shapes slightly differ each other,
depending on the criteria adopted either by the mid-point of the resistivity drop, the zero-resistivity, or by
thermalconductivity. We here consider the sharpest curve among them when several choices are available.

\subsubsection{$H\parallel b$: Reentrant SC}

\begin{figure}
\includegraphics[width=8cm]{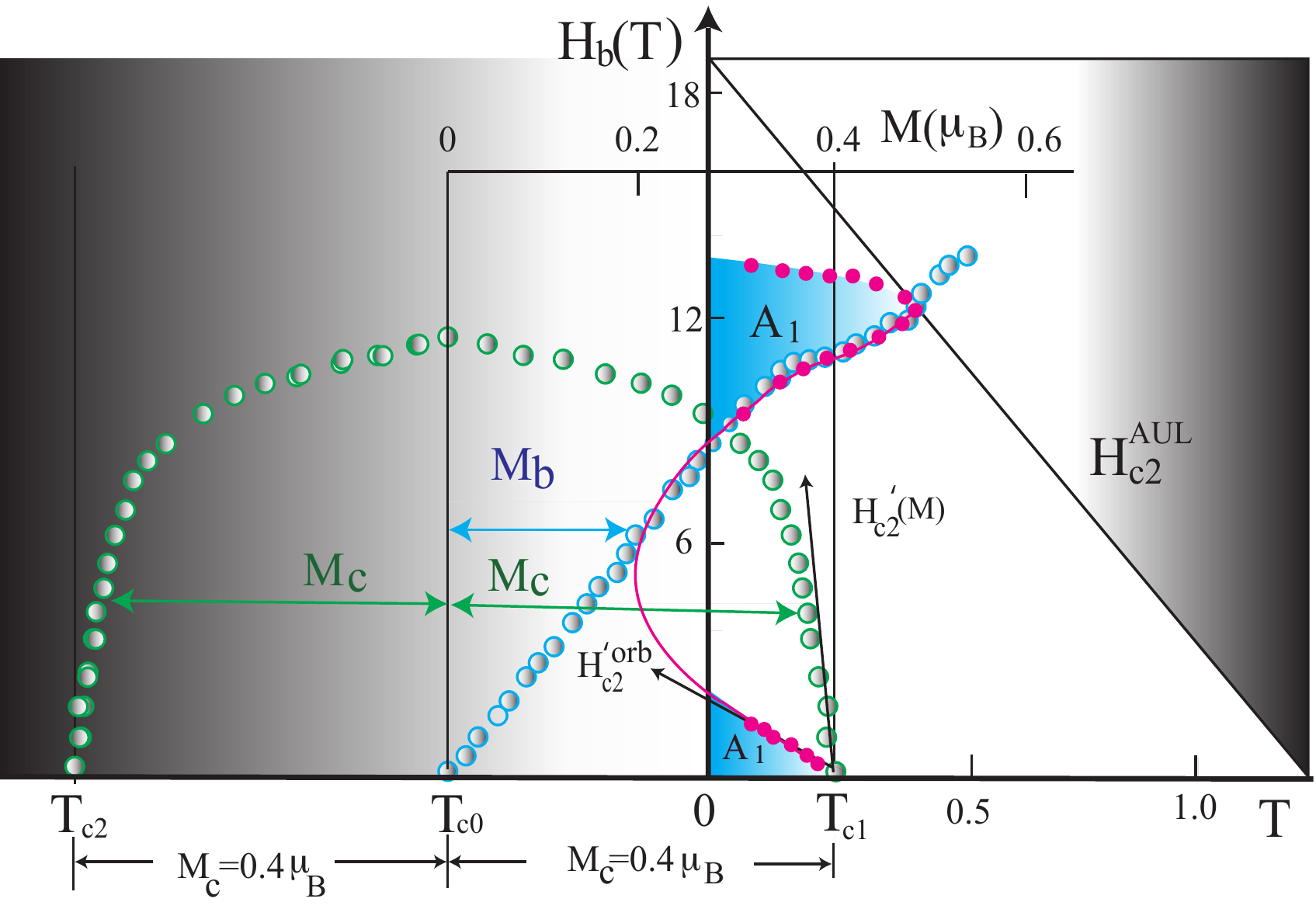}
\caption{(color online) 
The phase diagram for the $H_b$(T) versus $T$(K) plane.
$M_c(H_b)$  is estimated from the neutron scattering data in Ref.~[\onlinecite{levy}]
and $M_b(H_b)$ comes from the magnetization curve measured in Ref.~[\onlinecite{hardy}].
The red dots for $H_{\rm c2}$ are the experimental data points in Ref.~[\onlinecite{aokireview}].
The red continuous line indicates  $H_{\rm c2}$ which starts at $T_{c1}$  and is suppressed by the orbital depairing effect. 
It reappears again by following the formula $T_{c1}(H_b)=T_{c0}+\kappa M_b(H_b)$ near $H_R=11$T.
$H'^{\rm orb}_{\rm c2}$ ($H'_{\rm c2}(M)$) is the slope due to the orbital depairing ($T_{c1}(M_b)$).
}
\label{URhGe1-2}
\end{figure}

URhGe exhibits the ferromagnetic transition at $T_{\rm Cuire}=9.5$K where the magnetic 
easy axis is the $c$-axis and the FM moment $M_c=0.4\mu_B$. The superconducting transition is
at $T_{c}=0.4$K under the ferromagnetic state which is persisting to the lowest $T$.
When the field $H$ is applied parallel to the $b$-axis, the superconducting state reappears in a higher field region
while the low field SC phase disappears at $H_{\rm c2}\sim 2$T. This reentrant superconducting state (RSC) is explained 
in Fig.~\ref{URhGe1-2}, using the knowledge shown in Fig.~\ref{thetaphi}.

First we plot the magnetization curves for $M_c(H_b)$ and $M_b(H_b)$ in the $H$-$T$ plane by choosing the
$\kappa=2.0 {{\rm K}/\mu_{\rm B}}$ in Eq.~(\ref{tc3}) with $M_a$ replaced by $M_c$. 
$M_c(H_b)$ starts from $T_{c1}$ and $T_{c2}$ and decreases by increasing $H_b$
which acts to rotate the spontaneous ferromagnetic moment toward the $b$-axis as mentioned above.
Thus $T_{c1}(H_b)=T_{c0}+{\kappa} M_c(H_b)$ and $T_{c2}(H_b)=T_{c0}-{\kappa} M_c(H_b)$ 
decreases  and increases respectively with increasing $H_b$ according to Eq.~\eqref{tc3}.
The splitting $2{\kappa} M_c(H_b)$ between $T_{c1}(H_b)$ and $T_{c2}(H_b)$ 
diminishes and meets at the rotation field $H_{\rm R}$=12T
where the two transition temperatures are going to be degenerate.
$M_b(H_b)$ starting at $T_{c0}$ quickly increases there.
Thus as shown in Fig.~\ref{URhGe1-2},
$H_{\rm c2}$ starting at $T_{c1}$ disappears at a low field because the orbital depairing dominates
over the magnetization effect as explained above. Namely, since the decrease of $T_{c1}(H_b)$
is slow as a function of $H_b$, $H_{\rm c2}$ obeys the usual WHH curve, a situation similar to that shown in
Fig.~\ref{Hc2graph}(a). Here $|H^{\prime}_{\rm c2}(M)|\gg |H^{\prime orb}_{\rm c2}|$.

However, in the higher fields the upper transition temperature $T_{c1}(H_b)$ becomes

\begin{eqnarray}
T_{c1}(H_b)=T_{c0}+{\kappa} M_b(H_b)
\end{eqnarray}

\noindent
by rotating the $\bf d$-vector so that now it is perpendicular to the $b$-axis in order to grasp the magnetization $M_b(H_b)$.
This $\bf d$-vector rotation field corresponds to the field where

\begin{eqnarray}
T_{c1}(H_b)=T_{c0}+{\kappa} M_c(H_b)\simeq T_{c0}+{\kappa} M_b(H_b),
\end{eqnarray}

\noindent
namely,  the $M_c(H_b)$ vector projection onto the $b$-axis $M_c/\sqrt2\sim M_b(H_b)$ as understood from Fig.~\ref{Mc}.
Since $M_b(H_b)$ is strongly enhanced at and above $H_{\rm R}$, the A$_1$ phase reappears by following 
the magnetization curve $T_{c0}+{\kappa} M_b(H_b)$. It ultimately hits the $H^{\rm AUL}_{\rm c2}$
boundary. The RSC finally ceases to exist beyond this boundary. This corresponds to that in Fig.~\ref{Hc2graph}(c).
The existence of the $H^{\rm AUL}_{\rm c2}$ will be demonstrated later in Fig.~\ref{URhGePS} where we compile 
various $H_{\rm c2}$ data for URhGe, including those under hydrostatic pressure~\cite{miyake2} 
and uni-axial pressure~\cite{aoki-uni} along the
$b$-axis.

\subsubsection{$\theta$-rotation from $b$ to $c$-axis}

\begin{figure}
\includegraphics[width=6cm]{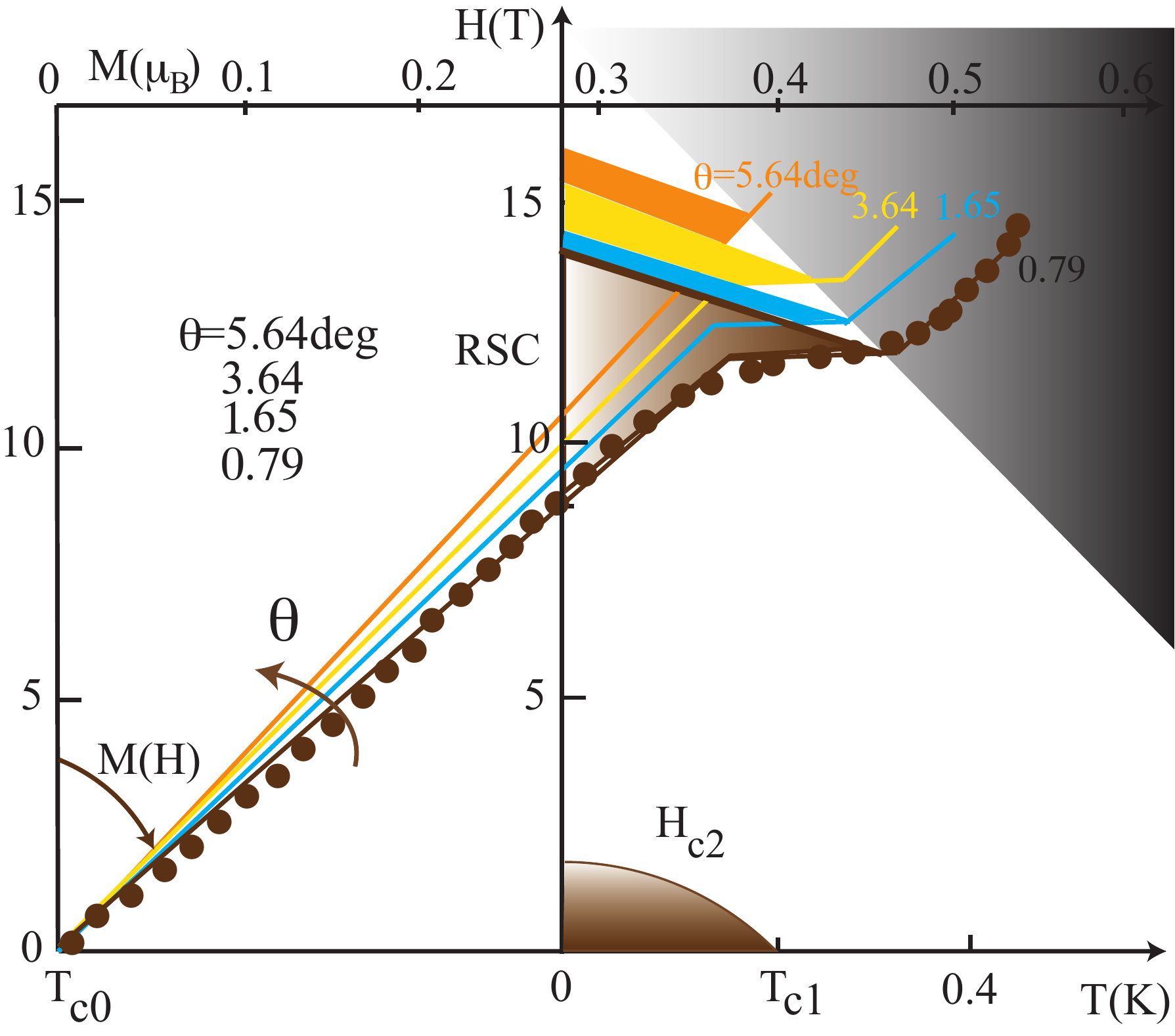}
\caption{(color online) 
Reentrant SC (Ref.~[\onlinecite{levy2}]) for various $\theta$ values measured from
the $b$-axis ($\theta=0$) toward the $c$-axis in URhGe.
As $\theta$ increases (0.79$^{\circ}$, 1.65$^{\circ}$, 3.64$^{\circ}$, and 5.64$^{\circ}$), 
the magnetization curves (far left scale) starting at $T_{c0}$ grows slowly, pushing
up the RSC regions to higher fields. The magnetization data are from Fig.~\ref{thetaphi}(a) for $\theta\neq 0$
and Ref.~[\onlinecite{hardy}] for $\theta=0$.
}
\label{thetaURhGe}
\end{figure}

When the direction of the magnetic field turns from the $b$-axis to the easy $c$-axis,
$T_R$ moves up to higher fields and disappears quickly around $\theta\sim5^{\circ}$ as shown in 
Fig.~\ref{thetaphi}(a).
According to those magnetization behaviors, we construct the $H_{\rm c2}$ phase diagram in 
Fig.~\ref{thetaURhGe}.
It is seen that the field-direction tilting away from the $b$-axis to the
$c$-axis results in the decrease of the magnetization $M_b(H)$, corresponding 
to the counter-clock wise changes of the magnetization curves in Fig.~\ref{thetaURhGe}. 
Thus the RSC region shifts to
higher fields with shrinking their areas and eventually disappears by entering the
$H_{\rm c2}^{\rm AUL}$ region.

\begin{figure}
\includegraphics[width=6cm]{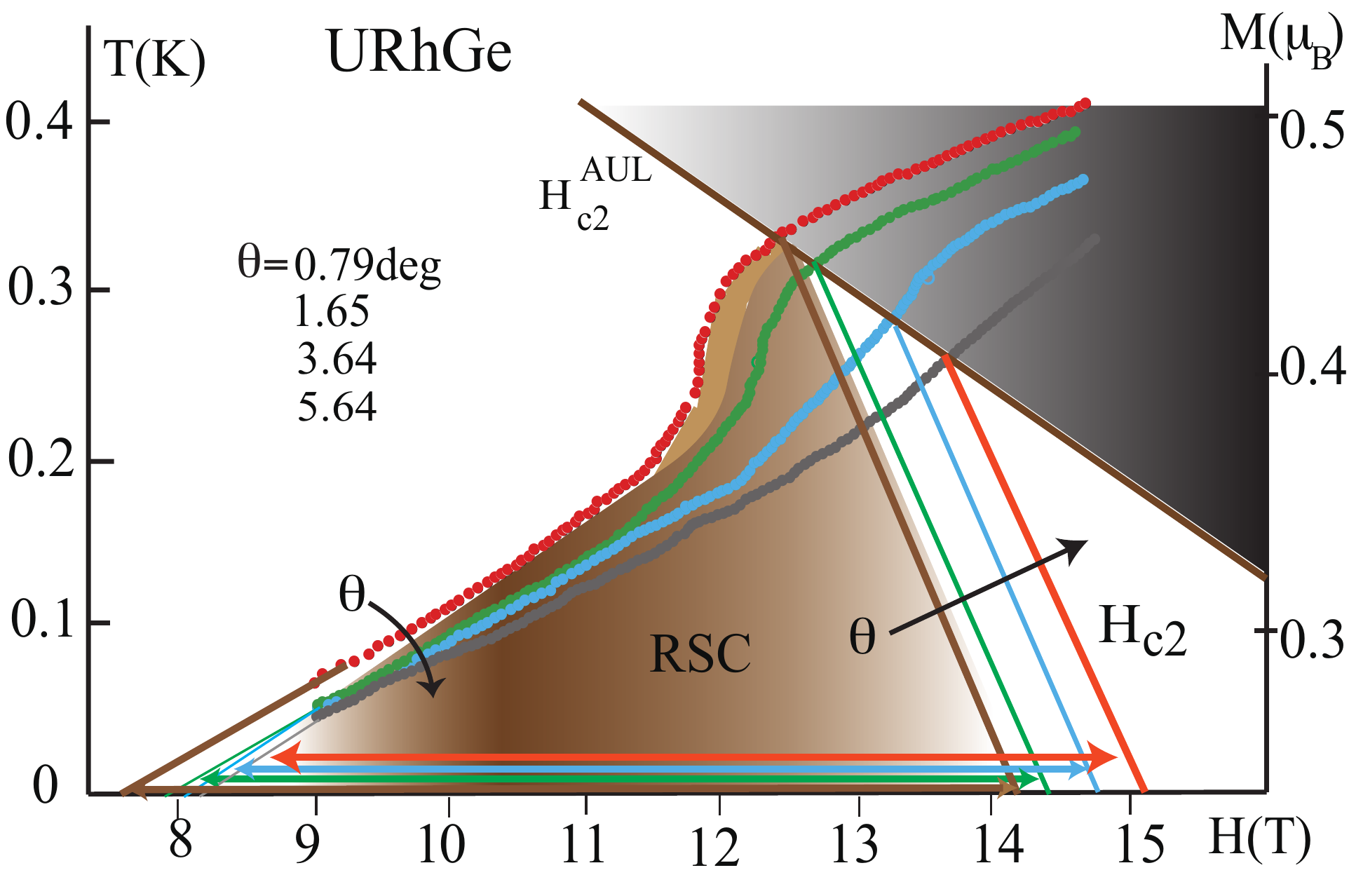}
\caption{(color online) 
Detailed RSC structures  (Ref.~[\onlinecite{levy2}])  in $T$-$H$ plane (left scale) are displayed.
The triangle areas in each  $\theta$ are RSC.
RSC moves right as $\theta$ increases.
The magnetization curve data (right scale) corrected as explained in Fig.~\ref{thetaphi}(a)
are originally from Ref.~[\onlinecite{nakamura}].
}
\label{Hc2-theta}
\end{figure}

The detailed phase diagram in the reentrant region is depicted in Fig.~\ref{Hc2-theta}
where the magnetization curves of $M_b(H)$ in Fig.~\ref{thetaphi}(a) are overlaid.
According to the present theory, $H_{\rm c2}$ follows faithfully $M_b(H)$ in
the high fields because the strong increase tendency of the magnetization $M_b(H)$
overcomes the orbital depression.
The characteristics of those phase diagrams are;  As $\theta$ increases,

\noindent
(1) The RSC moves up to further higher fields.

\noindent
 (2) As $H$  further increases,
within the small angles of $\theta$ up to $6^{\circ}\sim7^{\circ}$ the RSC fades 
out upon entering $H_{\rm c2}^{\rm AUL}$ region.

\noindent
Those characteristics (1) and (2) nicely match with the experimental observations.
The triangle-like shapes for RSC will be seen later in UTe$_2$ (see Fig.~\ref{theta35}).


\subsubsection{$\phi$-rotation from $b$ to $a$-axis}

\begin{figure}
\includegraphics[width=7cm]{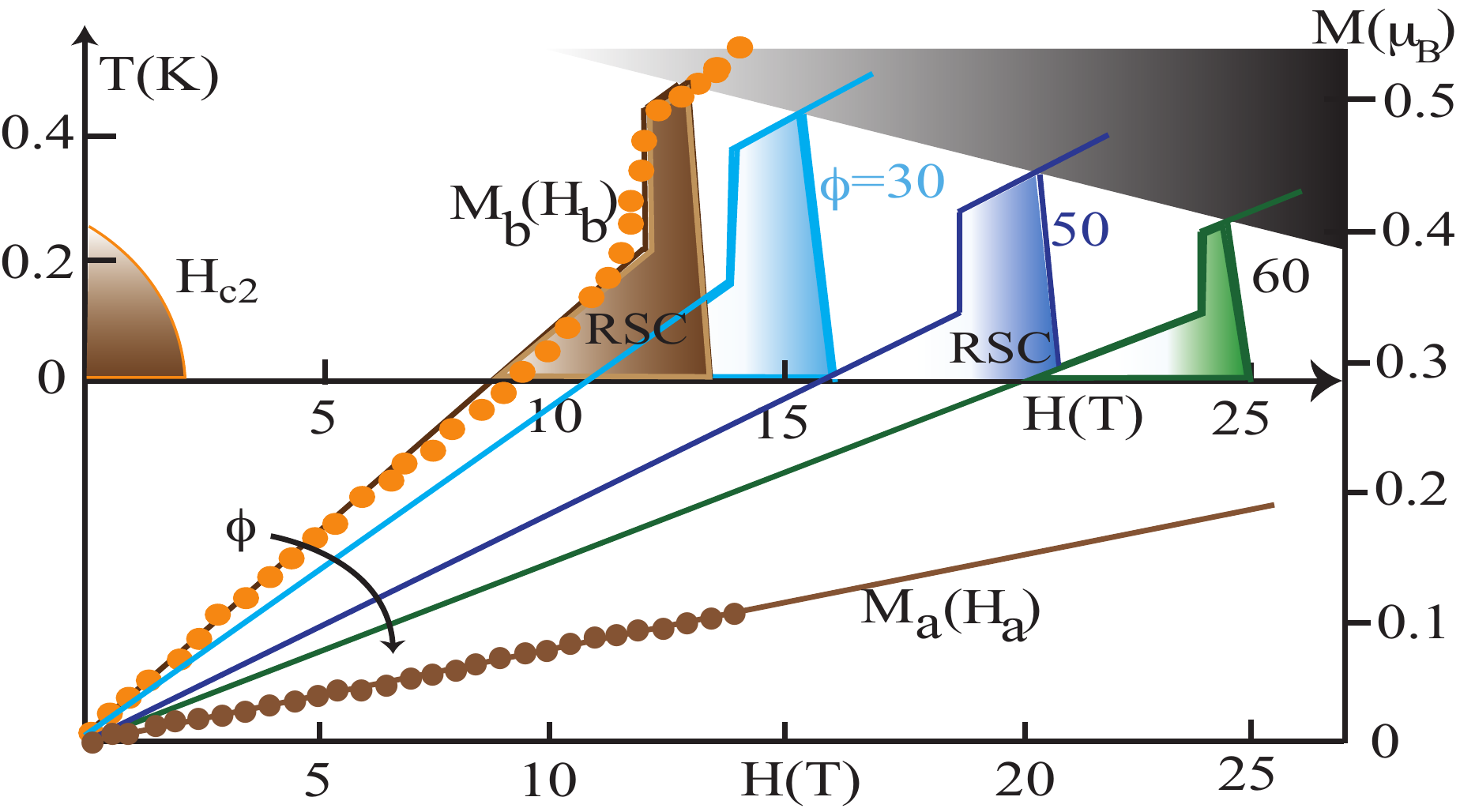}
\caption{(color online) 
RSC phase diagram in the $T$-$H$  plane for various fields rotated 
from the $b$-axis toward the $a$-axis by the angle $\phi$.
This is constructed by using the magnetization data (right scale) shown in 
Fig.~\ref{thetaphi}(b). When the magnetization hits the real axis $T>0$,
RSC appears in high field regions. The lower field $H_{\rm c2}$ is common for all $\phi$.
}
\label{phiRh}
\end{figure}

When the magnetic field direction turns to the other hard $a$-axis  from the $b$-axis
by the angle $\phi$, the expected magnetization curves are evaluated in Fig.~\ref{thetaphi}(b).
Using  those magnetization curves, we construct the $H_{\rm c2}$ phase diagrams for 
various $\phi$ values in Fig.~\ref{phiRh}. As the angle $\phi$ increases, the magnetization 
$M(H)$ decreases, corresponding to the clock-wise changes in Fig.~\ref{phiRh}
and the first order rotation field $H_R$ is pushed to higher fields
simply because of the projection effect onto the $b$-axis as mentioned in section IV-C. 
As a consequence, the RSC moves to
higher fields persisting up to higher angle $\phi$ until finally entering $H_{\rm c2}^{\rm AUL}$ region.
It is confirmed experimentally that it persists at least up to $H_{\rm c2}\sim25$T~\cite{levy}.
According to the present results, the RSC can exist still to higher fields.
This can be checked by experiments.

Here we notice an important fact that in order to explain the persistence of 
RSC as a function of $\phi$ up to higher fields, it is essential to use
the magnetization curves in Fig.~\ref{thetaphi}(b) where the magnetization
contains the component $M_a$ in addition to $M_b$.
It is clear that only $M_b$ fails to reproduce the RSC phase diagram.
This means that the $\bf d$-vector rotates so as to catch both components
$M_a$ and $M_b$, thus the $\bf d$-vector is always perpendicular to the
vectorial sum ${\bf M}_a+{\bf M}_b$. This is contrasted with the
$\theta$ rotation case mentioned above where the $\bf d$-vector is perpendicular to
${\bf M}_b$. This intriguing anisotropy in the $\bf d$-vector rotation relative to the
magnetic easy axis might be related to the underlying magnetism in
URhGe and/or the spin structure of the Cooper pair symmetry assumed as $SO(3)$ originally. 
This spin space anisotropy should be investigated in future.

\begin{figure}
\includegraphics[width=5cm]{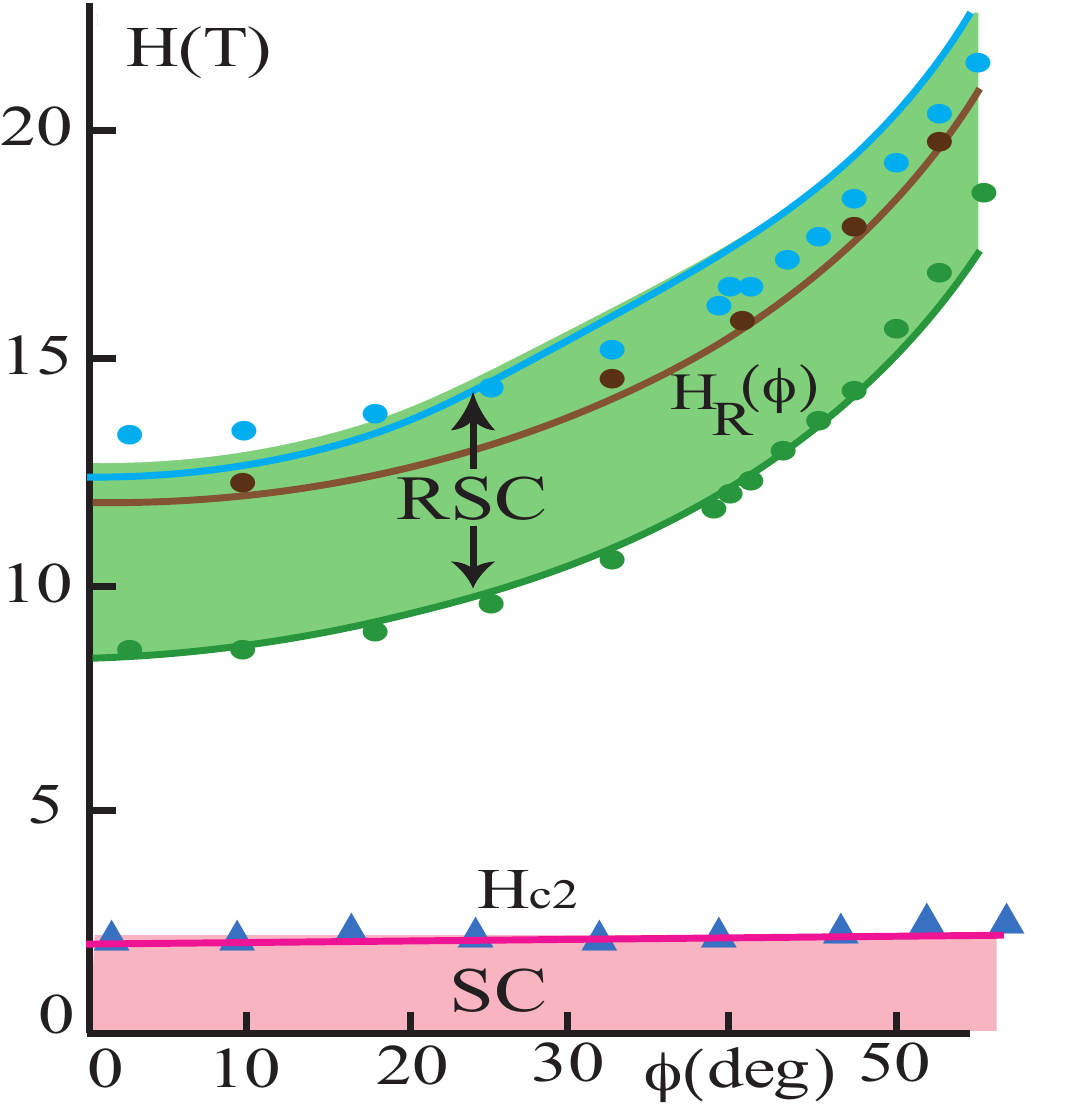}
\caption{(color online) 
Phase boundary of the reentrance SC (RSC) as a function
of the angle $\phi$ measured from the $b$-axis to the $a$-axis
constructed from Fig.~\ref{phiRh}.
The blue (green) line indicates the upper (lower) boundary of the RSC.
The brown line is the magnetization rotation field $H_R(\phi)$.
The dots are experimental data points by Ref.~[\onlinecite{levy2}].
The triangles denote the lower field $H_{\rm c2}$ which is almost independent
of $\phi$.
}
\label{RSC}
\end{figure}

In Fig.~\ref{RSC} we summarize the phase boundary of the RSC determined above.
The band of the RSC region is tightly associated with the $H_R(\phi)$ curves, which are proportional
to $H_R(\phi)\propto 1/\cos(\phi)$. This is contrasted with the lower field $H_{\rm c2}$
which is nearly independent of the angle $\phi$. The intrinsic $H_{\rm c2}$ anisotropy
is quite small in URhGe.
This means the importance of the magnetization rotation field $H_R(\phi)$, ensuring
the appearance of the RSC, and pointing to the simple mechanism for the origin of RSC.
It grossly follows the ${\bf M}_b$ projection onto the $b$-axis.
This is also true for the RSC in UTe$_2$, which will be explained shortly.
The physics is common.

\subsubsection{Pressure effects}

\begin{figure}
\includegraphics[width=7cm]{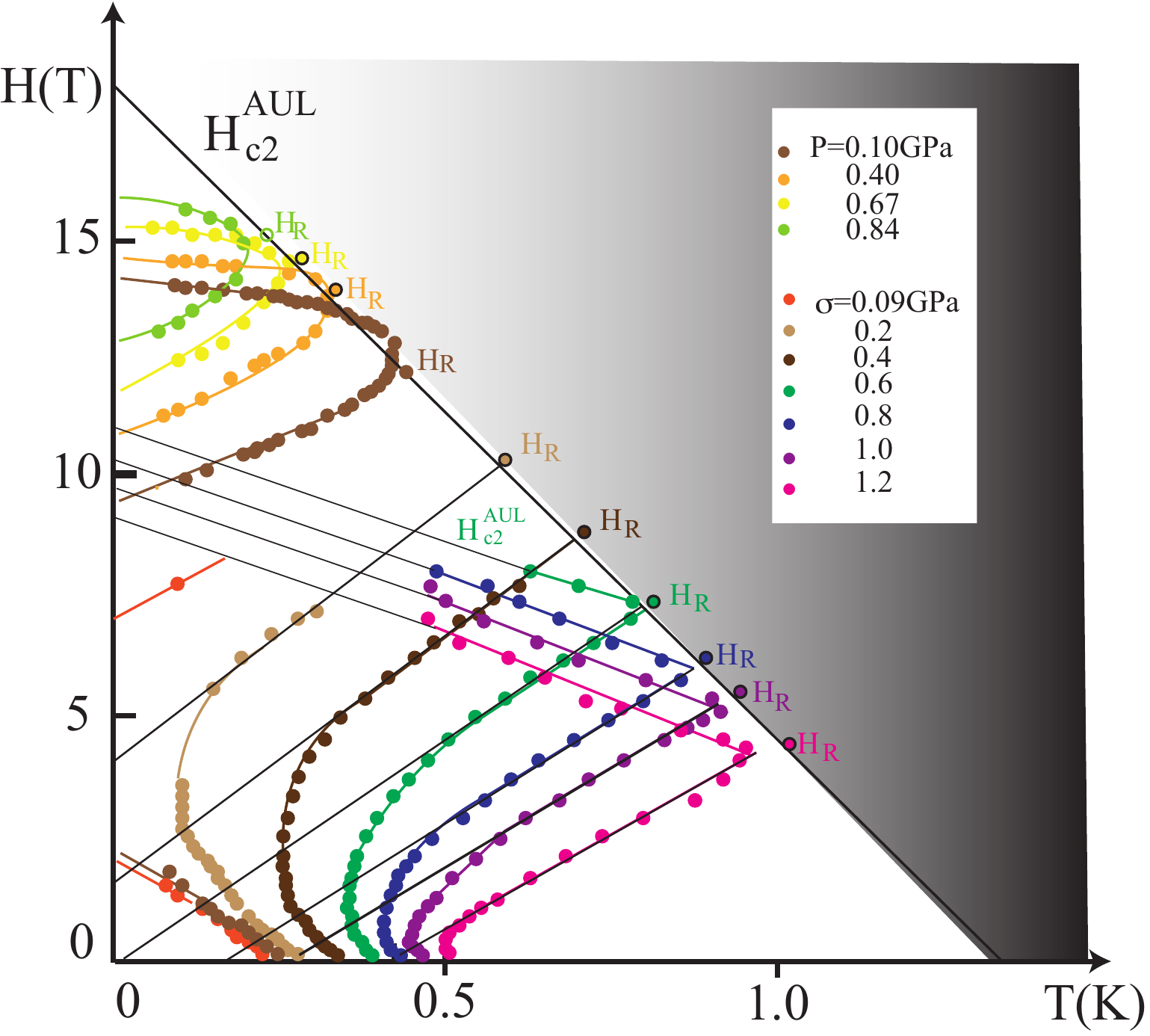}
\caption{(color online) 
Phase diagram for $H\parallel b$ taken under hydrostatic pressure (Ref.~[\onlinecite{miyake2}])
 and uni-axial pressure along the
$b$-axis (Ref.~[\onlinecite{aoki-uni}])
 on URhGe. All data at the rotation field $H_R$ line up along the $H^{\rm AUL}_{\rm c2}$ boundary,
 evidencing the existence of $H^{\rm AUL}_{\rm c2}$.
}
\label{URhGePS}
\end{figure}

Before starting out to analyze the experimental data taken under hydrostatic~\cite{miyake2}
and uni-axial pressure~\cite{aoki-uni} on URhGe, we summarize the relevant data for the
$H_{\rm c2}$ phase diagram with the field applied to the $b$-axis in Fig.~\ref{URhGePS}.
Here we list up the data under hydrostatic pressure and uni-axial pressure along the $b$-axis.

\noindent
(1) It is clear to see that all the $H_{\rm c2}$ are limited by the common boundary  $H^{\rm AUL}_{\rm c2}$.
Beyond $H^{\rm AUL}_{\rm c2}$ there exists no $H_{\rm c2}$ data.

\noindent
(2) It is also evident to see that the $H_R$ data points under pressure remarkably 
line up along the bottom of the boundary,
forming $H^{\rm AUL}_{\rm c2}$ as an envelop. 
In the following we utilize those experimental facts
and take into account those in investigating and reconstructing the $H_{\rm c2}$ phase diagrams.

\begin{figure}
\includegraphics[width=5cm]{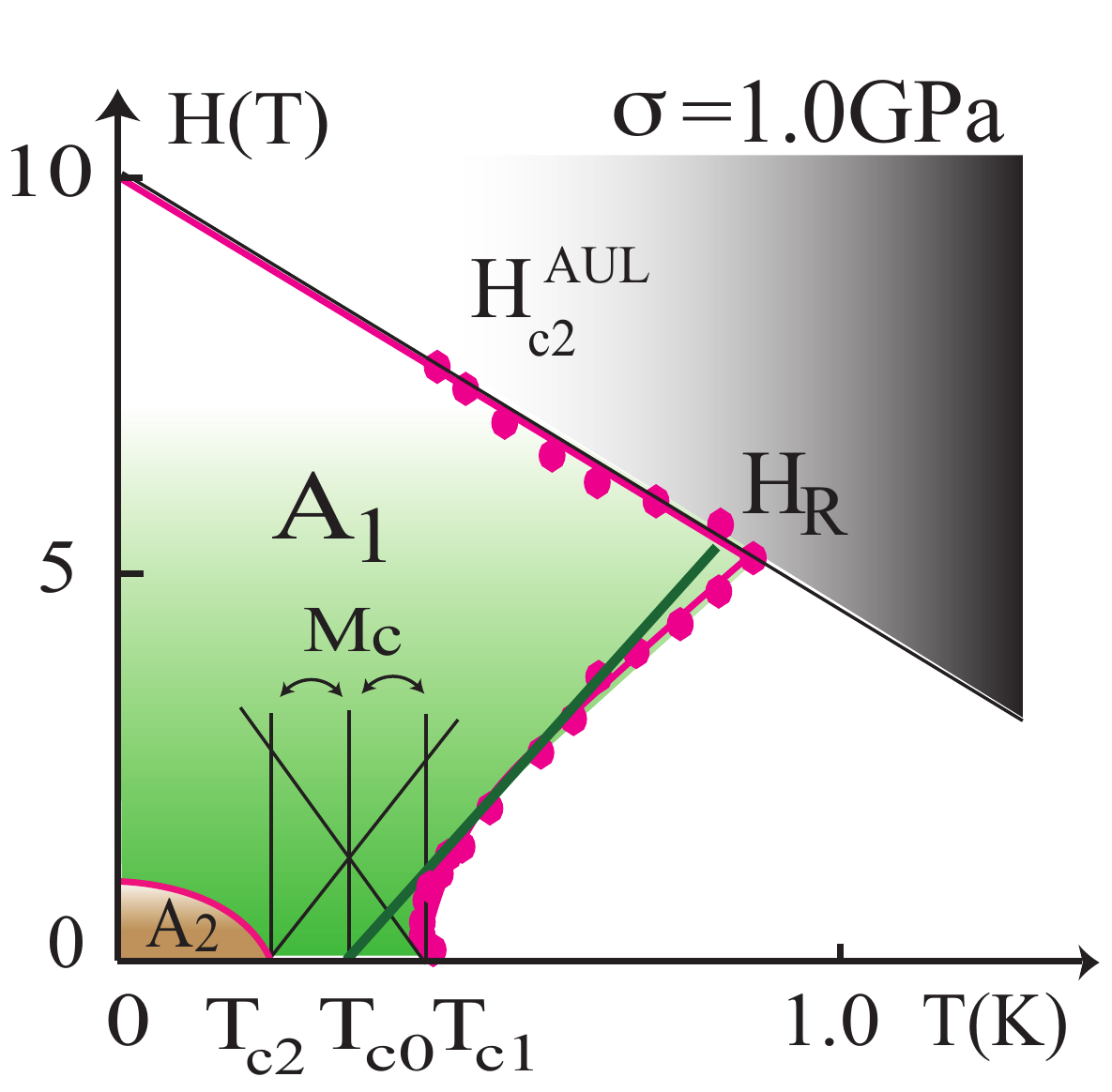}
\caption{(color online) 
Multiple phase diagram consisting of the $A_1$ and $A_2$ phases under uni-axial pressure $\sigma=1.0$GPa in
URhGe. The data points of $H_{\rm c2}\parallel b$ are taken from Ref.~[\onlinecite{aoki-uni}].
Two transitions at $T_{c1}$ and $T_{c2}$ separated by $2M_c$ are identified.
$H_R$ is the moment rotation field found experimentally~\cite{aoki-uni}. The green line indicates the magnetization curve of $M_b$
starting at $T_{c0}$.
}
\label{new1GPa}
\end{figure}

In Fig.~\ref{new1GPa} we show the $H_{\rm c2}$ data points taken when $H$ is applied along the $b$-axis
under uni-axial pressure $\sigma=1.0GPa$, which is listed in Fig.~\ref{URhGePS}. Those data are explained
in a similar way shown above. Here $H_{\rm c2}$ starting at $T_{c1}$  is strongly bent
due to the sharp $M_b(H_b)$ rise concomitant with the $\bf d$-vector rotation to catch $M_b(H_b)$
shown by the green line in Fig.~\ref{new1GPa}. Since $M_b(H_b)$ starts at the temperature $T_{c0}$
midway between $T_{c1}$ and $T_{c2}$ separated by $2M_c$, the second transition temperature $T_{c2}$ is found to locate there
where the $A_2$ phase begins developing while the remaining large region is occupied by the $A_1$ phase.
Now we see the multiple phases in this situation, which is absent under the ambient pressure in URhGe.
We can estimate the spontaneous moment $M_c$ under $\sigma$=1.0GPa
as $M_c=0.06\mu_B$ on the simple assumption that $\kappa$ is unchanged under the
uni-axial pressure.

\begin{figure}
\includegraphics[width=8.5cm]{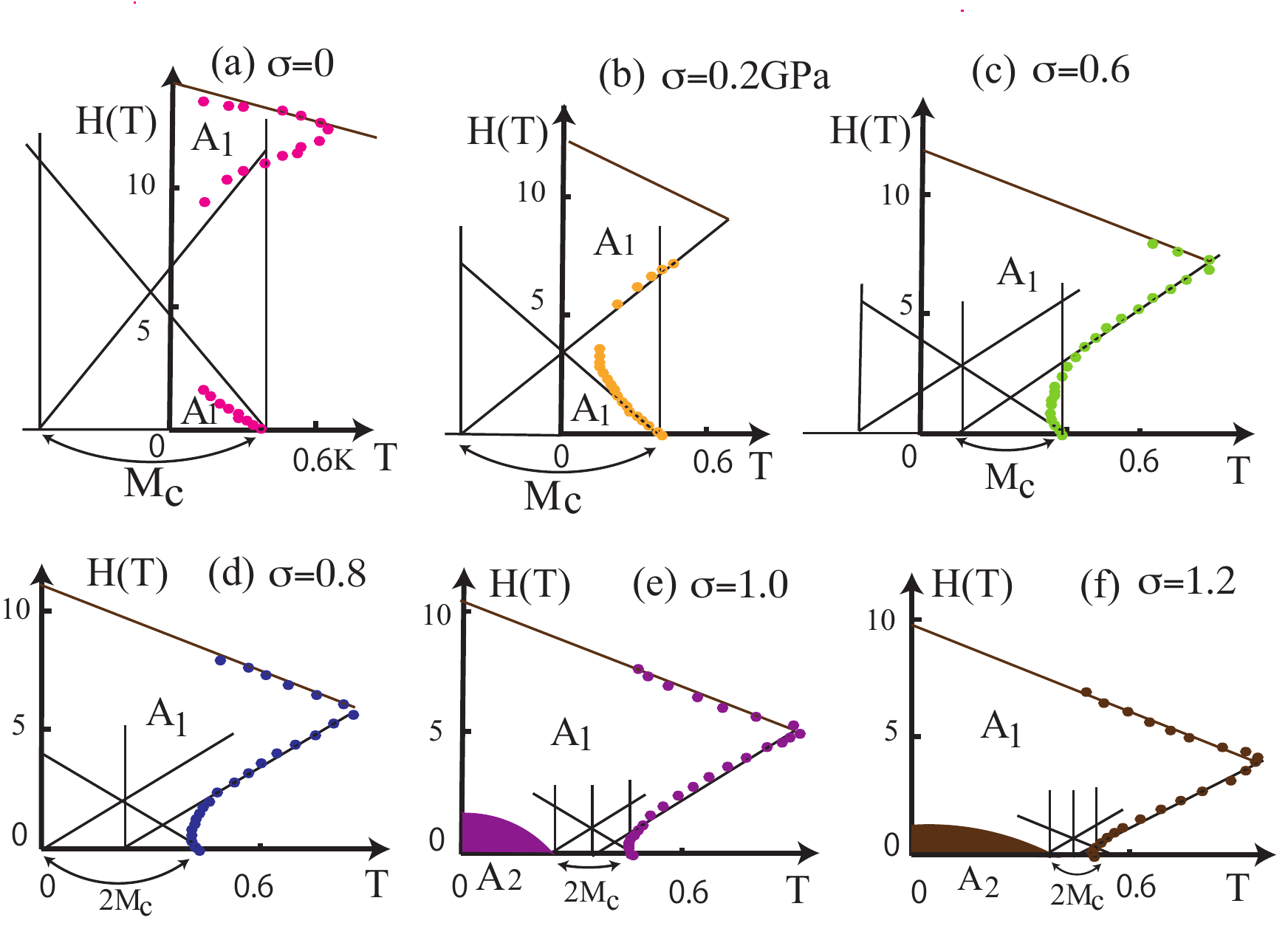}
\caption{(color online) 
Phase diagrams ($H\parallel b$) under uni-axial pressure, including the ambient pressure (a) in Fig.~\ref{URhGe1-2}
and $\sigma=1.0$GPa (e) in Fig.~\ref{new1GPa}. The data are from Ref.~[\onlinecite{aoki-uni}]. 
Continuous and systematic evolution of the multiple phase diagrams
with guide lines are seen. (a) $\sigma=0$GPa, (b) $\sigma=0.2$GPa, (c) $\sigma=0.5$GPa. (d) $\sigma=0.8$GPa,
(e) $\sigma=1.0$GPa, and (f) $\sigma=1.2$GPa.
}
\label{URhGeAll}
\end{figure}

\begin{figure}
\includegraphics[width=8cm]{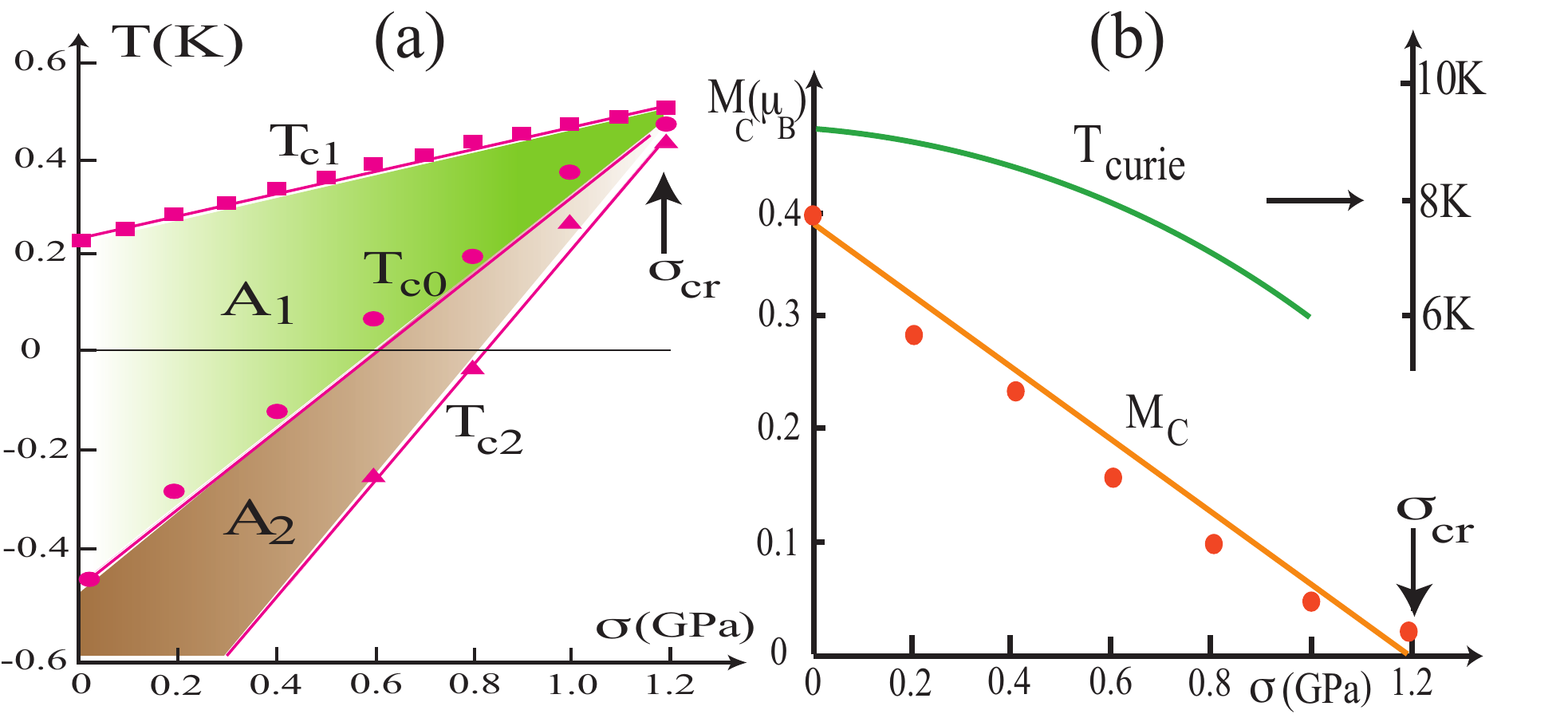}
\caption{(color online) 
(a) The resulting $T_{c1}$, $T_{c0}$ and $T_{c2}$ obtained from the analysis in Fig.~\ref{URhGeAll} are displayed.
The linear changes of those characteristic temperatures $T_{c1}$, $T_{c0}$ and $T_{c2}$ are found,
corresponding to the linear decrease in $M_c$. The second transition $T_{c2}$ begins appearing
above $\sigma>0.8$GPa where the double transitions are expected at $H$=0.
(b) The resulting $M_c$ change as a function of uni-axial pressure $\sigma$.
The observed Curie temperatures (Ref.~[\onlinecite{aoki-uni}]) are also shown.
It is consistent with the obtained decreasing tendency of $M_c$ as $\sigma$ increases.
}
\label{URhGesigma}
\end{figure}

We analyze the experimental data available under uni-axial pressure~\cite{aoki-uni}
displayed in Fig.~\ref{URhGeAll}. It is seen that the continuous and systematic evolution of the multiple phase diagrams under
uni-axial pressure. Namely, as uni-axial pressure $\sigma$ increases, three characteristic temperatures
$T_{c1}$, $T_{c0}$ and $T_{c2}$ shifts together to higher temperatures. $T_{c2}$ appears at a finite temperature ($T>0$)
around $\sigma\sim0.8$GPa, keeping to move up with increasing further $\sigma$.
The separation of $T_{c1}$ and $T_{c2}$ becomes narrow because the spontaneous moment $M_c$ gets diminished,
corresponding to the observed Curie temperature decrease 
under uni-axial pressure~\cite{aoki-uni} (see Fig.~\ref{URhGesigma}(b)).

We show the changes of three temperatures $T_{c1}$, $T_{c0}$ and $T_{c2}$ assigned thus in Fig.~\ref{URhGesigma}(a).
The separation between $T_{c1}$ and $T_{c2}$ determined by $M_c$  diminishes simply because 
$M_c$ decreases as $\sigma$ increases. This results in $T_{c2}>0$ appearing  above $\sigma>0.8$GPa,
where the double transitions at $H$=0 should be observed. It is remarkable to see that upon approaching 
$\sigma=1.2$GPa from below, all the transition temperatures are converging toward $\sigma_{\rm cr}=1.2$GPa. 
This means that above this pressure,
the genuine symmetric $A$ phase is realized because the symmetry breaking parameter $M_c$ vanishes
where the spin symmetry of the pair function restores $SO(3)$ full symmetry, a situation 
similar to that shown in Fig.~\ref{ProtoA0} (also see Fig.~\ref{UTe2PT} later).
At the critical pressure $\sigma_{cr}$=1.2GPa the pairing state is analogous to superfluid $^3$He-$A$ phase.

The resulting analysis of the spontaneous moment $M_c$ is shown in Fig.~\ref{URhGesigma}(b),
revealing a monotonous decrease as $\sigma$ increases. This tendency is matched with the 
lowering of the Curie temperature, which is observed experimentally~\cite{aoki-uni}.
It is interesting to see the linear changes of $T_{c1}$, $T_{c0}$, $T_{c2}$, and $M_c$ near the critical uni-axial pressure 
$\sigma_{\rm cr}=1.2$GPa. This linear relationship is similar to those in UTe$_2$ under hydrostatic 
pressure around the critical pressure $P_{cr}$=0.2GPa (see  Fig.~\ref{UTe2PT} later).

\subsection{UCoGe}

UCoGe is another ferromagnetic superconductor worth checking our theory in the same framework for URhGe.
Major differences from URhGe in the previous section lie in the fact that 

\noindent
(1) The small spontaneous moment $M_c=0.06\mu_B$.

\noindent
(2) The field induced moments of $M_b$ and $M_a$ in the hard axes are comparable in magnitude
as shown in Fig.~\ref{Mall}(a).

\noindent
(3) The magnetization rotation field $H_R\sim 45$T is far above $H_{\rm c2}$.
Those are contrasted with URhGe with the distinctive induced moment for $M_b$ that ultimately leads to the RSC.
However, $H_{\rm CR}$ is situated at low fields $6\sim8$T in UCoGe.

\subsubsection{$H\parallel b$: S-shaped $H_{\rm c2}$ and multiple phases}

\begin{figure}
\includegraphics[width=6cm]{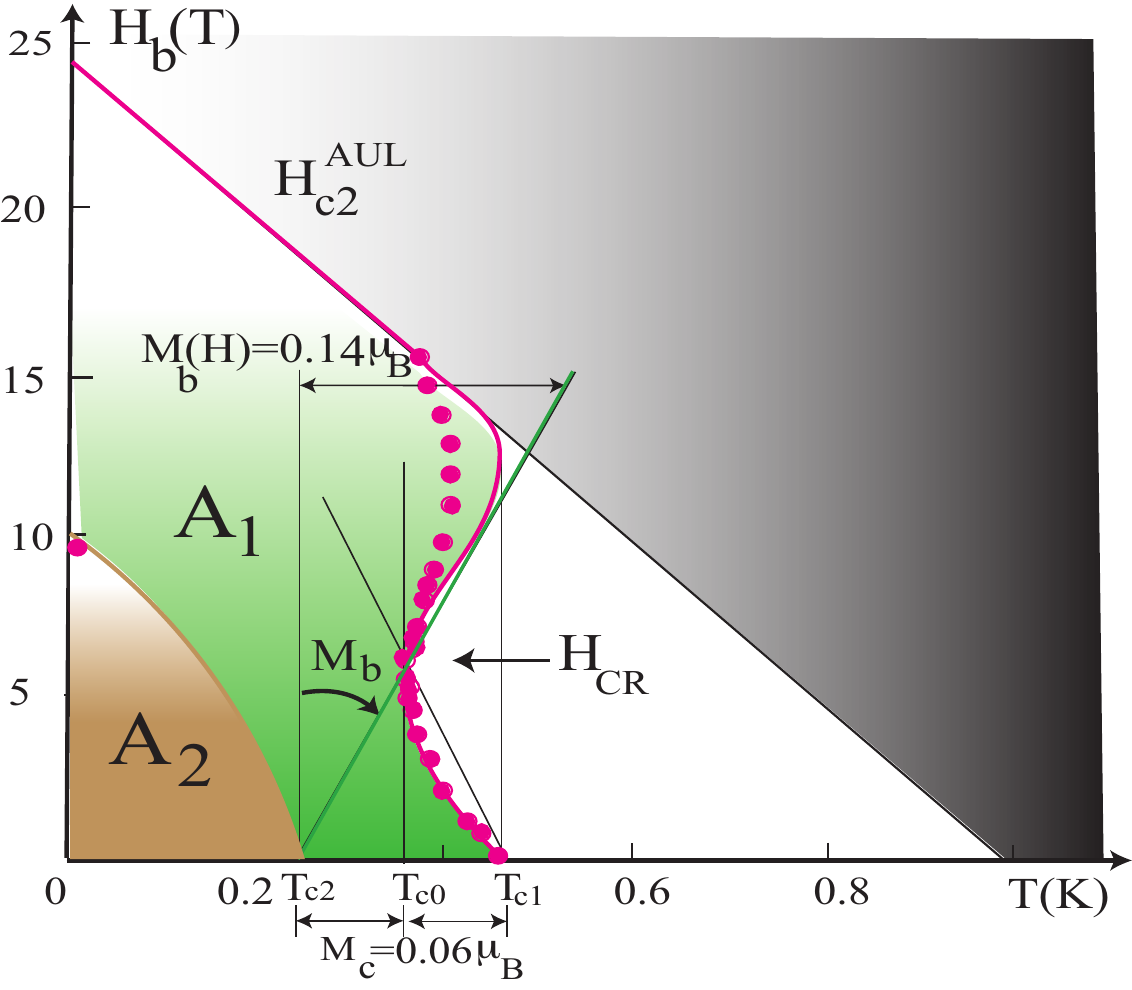}
\caption{(color online) 
The S-shaped phase diagram for UCoGe in $H\parallel b$.
$H^b_{\rm c2}$ starts at $T_{c1}$ is initially depressed
by the orbital depairing. At around the crossing field $H_{\rm CR}$
it turns toward higher $T$ due to the $\bf d$-vector rotation to catch $M_b(H_b)$
denoted by the green line, forming the S-shape. At further high fields
after hitting $H^{\rm AUL}_{\rm c2}$, $H^b_{\rm c2}$ follows it.
The experimental data points come from [\onlinecite{aokiS}]
and the point at $T$=0 and 10T from [\onlinecite{wu}].
}
\label{UCoGe1}
\end{figure}

In Fig.~\ref{UCoGe1} we show the result for the phase diagram in $H\parallel b$,
assuming that $\kappa=1.8{K\over \mu_B}$.
The two transition temperatures $T_{c1}$ and $T_{c2}$ are split by $M_c=0.06\mu_B$.
Under the applied field $H_b$, the spontaneous moment $M_c (H_b)$ decreases. $T_{c1}$ and $T_{c2}$
approach each other to meet at $H^b_{\rm CR}\sim6$T. Before meeting there, the upper $T_{c1}(H_b)$
increases and catches the magnetization $M_b(H_b)$ by rotating the $\bf d$-vector direction from the
$c$-perpendicular direction to the $b$-perpendicular direction. This results in an S-shaped $H_{\rm c2}$ curve
which eventually reaches  $H^{\rm AUL}_{\rm c2}$, giving the extrapolated $H^b_{\rm c2}\sim25$T.
We notice here that the initial slope of $H^b_{\rm c2}$ is small,
extrapolated to $H^b_{\rm c2}$ less than a few T, which is comparable to $H^c_{\rm c2}\sim0.5$T.
This means that the intrinsic $H_{\rm c2}$ anisotropy is within the range of the usual effective mass anisotropy.
The same nearly isotropic  $H_{\rm c2}$ behavior was just emphasized
in URhGe (see Fig.~\ref{RSC}).
The superficial $H_{\rm c2}$ anisotropy with the order of $H^b_{\rm c2}/H^c_{\rm c2}$=25T/0.5T$\sim50$ is an artifact
due to ignoring the origin of the S-shaped $H^b_{\rm c2}$.
This is often pointed out as one of the major mysteries in UCoGe~\cite{aokireview}.

It is important to notice that because we identify $T_{c2}=0.2$K there must exist the phase boundary
of $A_1$ and $A_2$ phases. According to thermal-conductivity measurement in  Ref.~[\onlinecite{wu}] as a function of $H_b$,
there indeed exists an anomalous thermal-conductivity jump at 10T and low $T$  indicated 
as the red dot on the $H$-axis in Fig.~\ref{UCoGe1}.
This nicely matches our identification of the $A_2$ phase boundary line, a situation similar to the characteristics in
Fig.~\ref{Hc2graph}(c) and Fig.~\ref{protoHc2b}(b).
This assignment is consistent with the $H^c_{\rm c2}$ phase diagram as shown shortly.

\begin{figure}
\includegraphics[width=8.5cm]{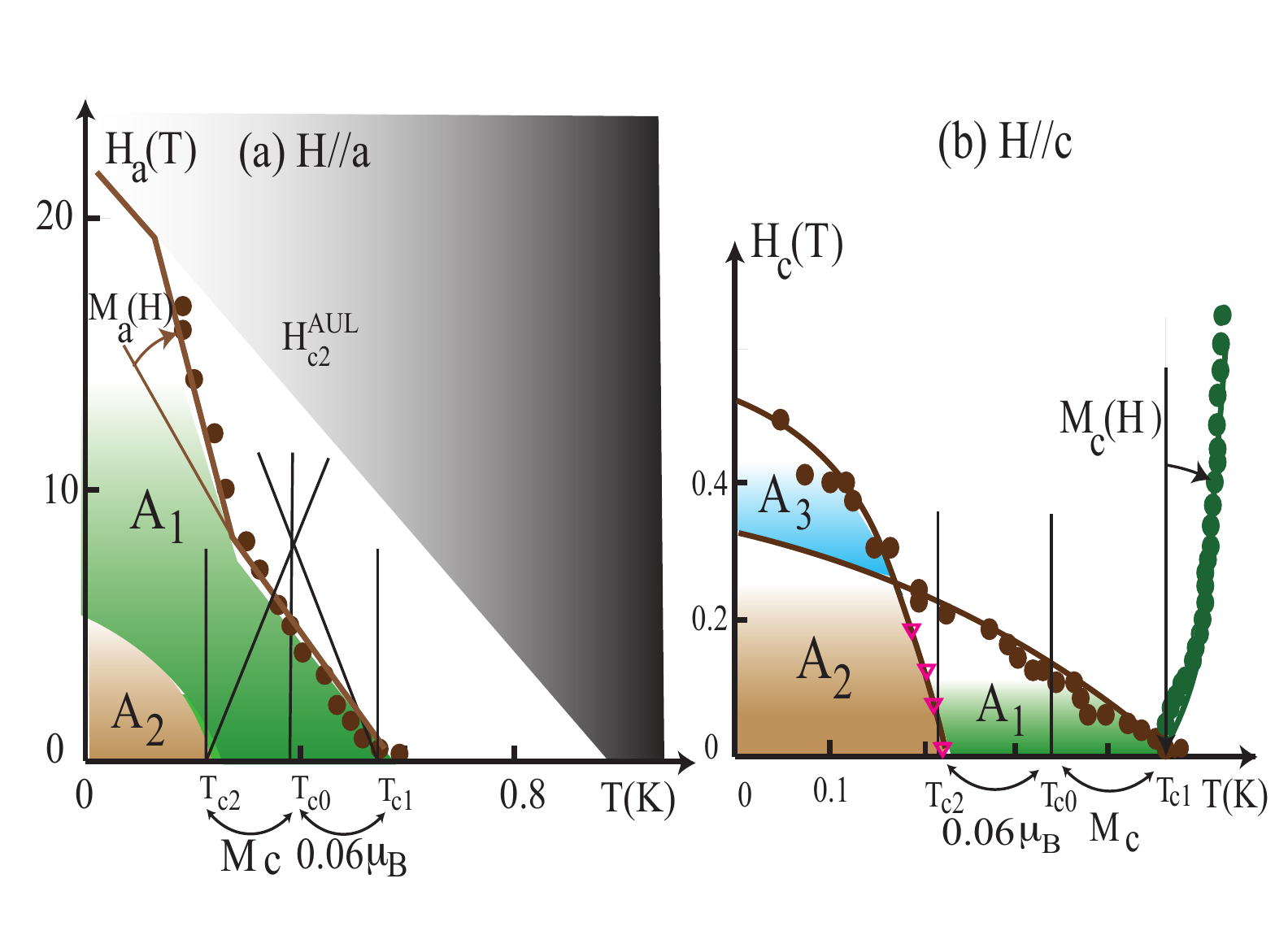}
\caption{(color online) 
(a) Weaken S-shaped $H^a_{\rm c2}$ for $H\parallel a$ in UCoGe because $H_{\rm \rm CR}$ moves up compared to 
$H^b_{\rm c2}$ case shown in Fig.~\ref{UCoGe1}.  The data are from Ref.~[\onlinecite{aokiS}].
(b)  $H^c_{\rm c2}$ for $H\parallel c$ in UCoGe. The data~[\onlinecite{wu}] clearly show the anomaly around 0.3T,
indicating the multiple phases identified as $A_1$, $A_2$, and $A_3$. The magnetization curve of $M_c(H_c)$
is displayed as the green dots, showing the weak rise in this scale. 
Both $H^c_{\rm c2}$ starting at $T_{c1}$ and  $T_{c1}$ are thus dominated by the
orbital depairing without help of the magnetization. The four points denoted by the red triangles are read off from the thermal-conductivity
anomalies [\onlinecite{taupin}].
}
\label{UCoGe-ac}
\end{figure}

\subsubsection{$H\parallel a$}

As already shown in Fig.~\ref{Mall}(a), the magnetization curves of $M_b$ and $M_a$
is quite similar. The crossing field $H^{i}_{\rm \rm CR}$ ($i=a$ and $b$) at which $M_b(H_b)$ and $M_a(H_a)$ reach $M_c=0.06\mu_B$
is seen to be $H^{a}_{\rm \rm CR}\sim8$T and $ H^{b}_{\rm \rm CR}\sim6$T.
Thus $H^c_{\rm c2}$ curve is anticipated to be similar too. Indeed the result is shown in Fig.~\ref{UCoGe-ac}(a).
Even though the S-shaped $H^b_{\rm c2}$ is weaken, it is still seen a weak anomaly at around
$H^{a}_{\rm \rm CR}\sim8$T which is a signature that $T_{c1}(M)$ in Eq.~(\ref{tc3}) catches $M_a(H_a)$ by rotating the
$\bf d$-vector whose direction is perpendicular to the $c$-axis. It is now perpendicular to the $a$-axis.
We also point out that the $A_1$ and $A_2$ phase diagram is essentially the same as in $H\parallel b$
and the extrapolated $H^a_{\rm c2}\sim$22T is comparable to $H^b_{\rm c2}\sim$25T.

\subsubsection{$H\parallel c$ and multiple phases}

We display the analysis for the phase diagram in $H\parallel c$ in Fig.~\ref{UCoGe-ac}(b).
The existing experimental data clearly indicate that $H^c_{\rm c2}$ 
consists of the two parts where the $H^c_{\rm c2}$ enhancement is visible at low $T$ and high $H$. 
Thus the phase diagram is divided into the three phases,
$A_1$, $A_2$ and $A_3$ where $A_3$ is genuine spin down-down pair while $A_2$
is a mixture of up-up and down-down pairs, or a distorted $A$ phase with different population of 
the two spin pairs.

As indicated in Fig.~\ref{UCoGe-ac}(b) as the green dots, the magnetization curve of $M_c(H_c)$ is weakly increasing
in this scale. Thus the slope at $T_{c1}$ is exclusively governed by the orbital depairing, implying that this comes from
the effective mass along the $c$-axis. As mentioned above, the anisotropy of the initial slopes in $H_{\rm c2}$
at $T_{c1}$ is determined by their effective mass anisotropy.

\subsubsection{Rotation $\phi$ from the $b$-axis toward the $a$-axis}

\begin{figure}
\includegraphics[width=5cm]{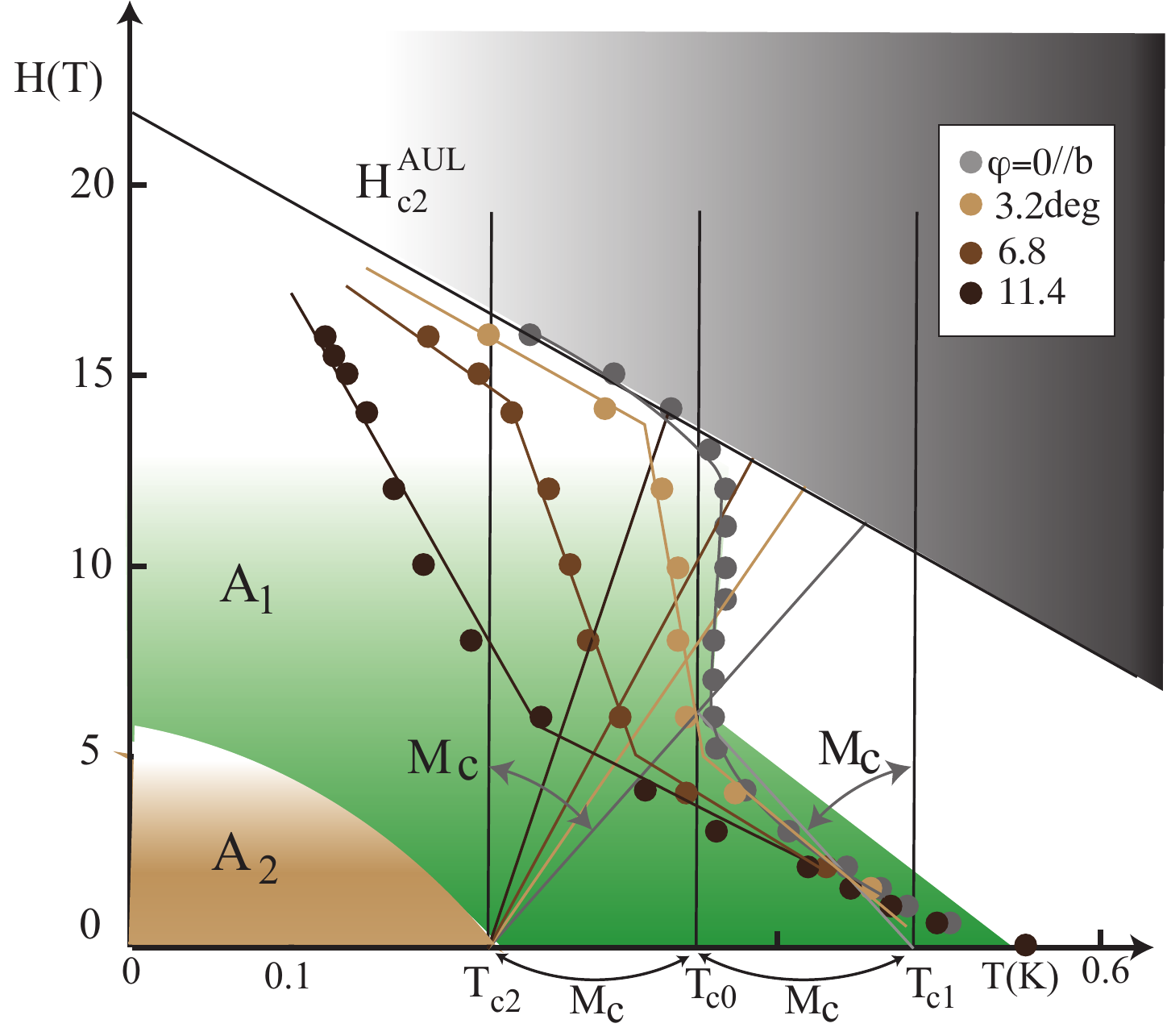}
\caption{(color online) 
$H_{\rm c2}(\phi)$ for $\phi=0$$^{\circ}$, 3.2$^{\circ}$, 6.8$^{\circ}$ and 11.4$^{\circ}$ from the $b$-axis toward the
$a$-axis in UCoGe. The data are from Ref.~[\onlinecite{aokiS}].
As $\phi$ increases, $M_c$ grows slowly as a function of $H$ (the counter-clock wise rotation of the $M_c$ curves),
pushing up $H_{\rm CR}$ to higher fields. This results in the decreases of $H_{\rm c2}(\phi)$
because the orbital suppression  becomes dominant. The  enhanced $H_{\rm c2}$ becomes diminished
as $\phi$ increases.
}
\label{UCoGe2}
\end{figure}

Finally we touch upon the case of the field rotation from the $b$-axis toward the $a$-axis by $\phi$
as shown in Fig.~\ref{UCoGe2}.
As $H$ is turned from the $b$-axis toward the other hard  $a$-axis, the crossing field $H_{\rm CR}$ increases
as shown in Fig.~\ref{Mall}(a). Since $M_c(H)$ becomes slowly increasing as $H$ increases,
the orbital depression gets stronger and flattens the initial slopes of $H_{\rm c2}(\phi)$ at $T_{c1}$,
eventually approaching $H^a_{\rm c2}$ as shown in Fig.~\ref{UCoGe-ac}(a).  This is already realized in 
$\phi=11.4^{\circ}$ case seen from it.
It should be pointed out again that those initial slopes at $T_{c1}$  for those $\phi$ values
only slightly change, implying that the initial slope is determined by the effective masses, namely
the orientational dependent Fermi velocities.

So far we assumed that $\kappa=1.8{K/\mu_B}$ under the condition of the existence of
the second transition $T_{c2}=0.2$K. But we are warned that if those suggestive signatures of 
the second phase $A_2$ coming from thermal-conductivity measurements~\cite{wu,taupin} may be an 
artifact, then the forgoing arguments go through and are almost unchanged by taking
$\kappa=3.6{K/ \mu_B}$ without the A$_2$ phase. Namely we are still in ambiguous situations to finally pin down 
the system parameters. Therefore, it is urgent to confirm or refute the existence of the second transition
in order to go further from here.



\subsection{UTe${_2}$}

To coherently explain a variety of physical properties of superconducting state in UTe${_2}$ accumulated experimentally
in the same context of the other compounds, URhGe and UCoGe, 
we need a basic assumption that the ferromagnetic fluctuations are slow enough compared to
the electron motion of the conduction electrons, which condense at $T_c$.
The slow FM fluctuation moments characterized by the non-vanishing square root-mean averaged value 
$\sqrt{\langle(\delta M_a)^2\rangle}$
over time and space $\langle\cdots\rangle$
are assumed to be able to break the spin symmetry SO(3) of the
Copper pairs. In the following we denote this spontaneous and instantaneous 
FM moment simply $M_a$=$\sqrt{\langle(\delta M_a)^2\rangle}$,
whose magnitude is adjusted in order to best reproduce the $H_{\rm c2}$ phase diagram as we will see next.

\subsubsection{$H\parallel b$-axis}

\begin{figure}
\includegraphics[width=7cm]{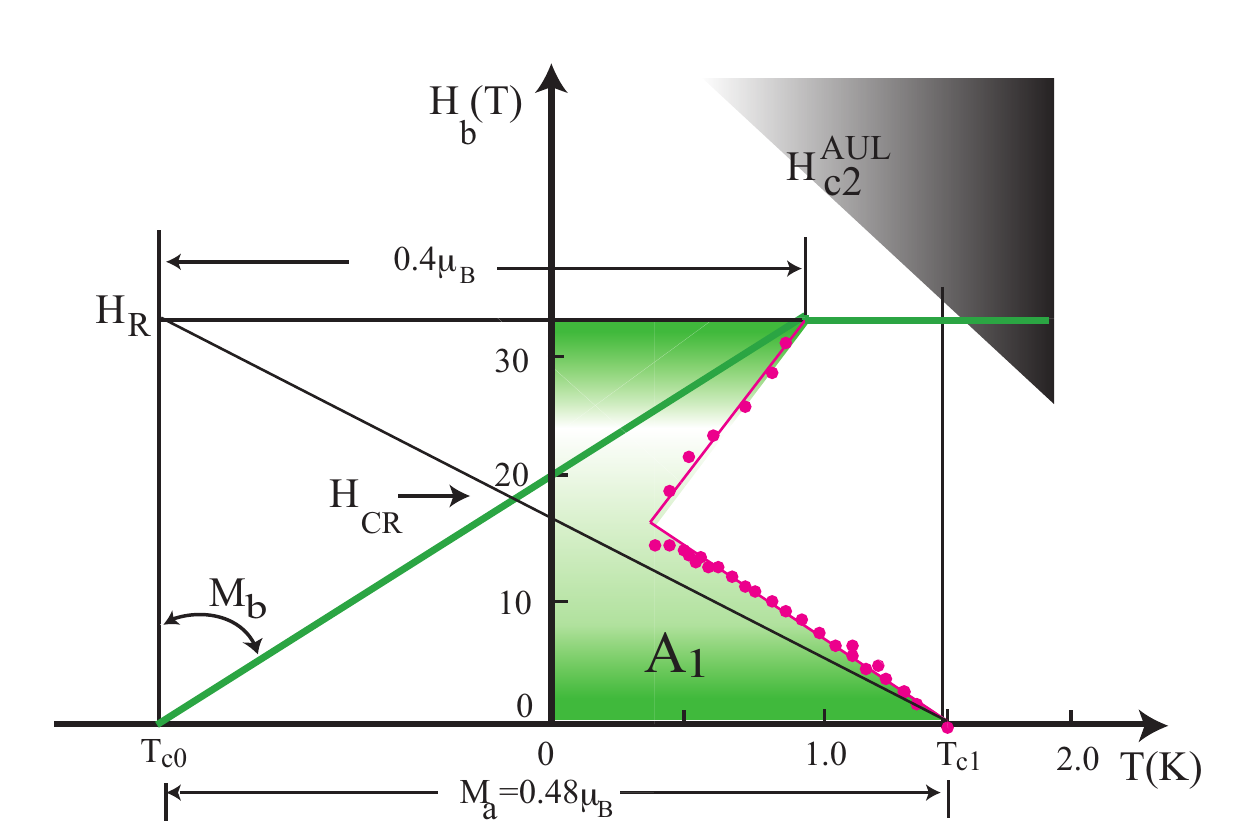}
\caption{(color online) 
The L-shaped $H^b_{\rm c2}$ observed in [\onlinecite{knebel}] is shown (red dots).
$H^b_{\rm c2}$ starting at $T_{c1}$ follows the orbital suppression plus the
$M_a$ depression by $H_b$ toward $H_{\rm \rm CR}$. When it approaches the strong increasing $M_b(H_b)$,
the $\bf d$-vector rotates and catches $M_b(H_b)$ to grow. This forms the upper part of the L-shape.
In further high fields $H^b_{\rm c2}$ reaches $H_R$=32T and disappears there by hitting $H^{\rm AUL}_{\rm c2}$.
The green curve denotes the magnetization curve $M_b(H_b)$ shown in the inset of Fig.~\ref{Mall}(b)[\onlinecite{miyake}].
}
\label{UTe2}
\end{figure}

We follow the same method mentioned above for URhGe and UCoGe 
to understand the observed L-shaped $H^b_{\rm c2}$ applied to the magnetic hard $b$-axis.
Here we assume that  $M_a=0.48\mu_B$ and $\kappa=6.9{K/ \mu_B}$.
As seen from Fig.~\ref{UTe2}, $H^b_{\rm c2}$ starts from $T_{c1}$=1.6K, following $M_a(H_b)$ which decreases
with increasing $H_b$ toward $H_{\rm \rm CR}$. $H_{\rm \rm CR}$  is roughly estimated from the magnetization curves
shown in the inset of Fig.~\ref{Mall}(b) as around 20T. Above $H_b>H_{\rm CR}$ the $\bf d$-vector rotates
in order to catch the magnetization $M_b(H_b)$,
which strongly increases from $T_{c0}$. $H^b_{\rm c2}$  begins following it to grow and forms the upper part of the L-shape.
It eventually reaches $H_R$=32T where the first order transition occurs. As shown in the inset of Fig.~\ref{Mall}(b)
the magnetization jump at $H_R$ amounts to 0.6$\mu_B$
known experimentally~\cite{miyake}.
The reached magnetization (the horizontal green line) is deep outside $H^{\rm AUL}_{\rm c2}$ shown 
by dark black colored region in Fig.~\ref{UTe2}.
Therefore $H^b_{\rm c2}$ simply stops when it hits the $H_R$ line.
Those features nicely reproduce the experimental characteristics shown in Fig.~\ref{UTe2}.

\subsubsection{$\phi$ rotation from the $b$-axis toward the $a$-axis}

\begin{figure}
\includegraphics[width=7cm]{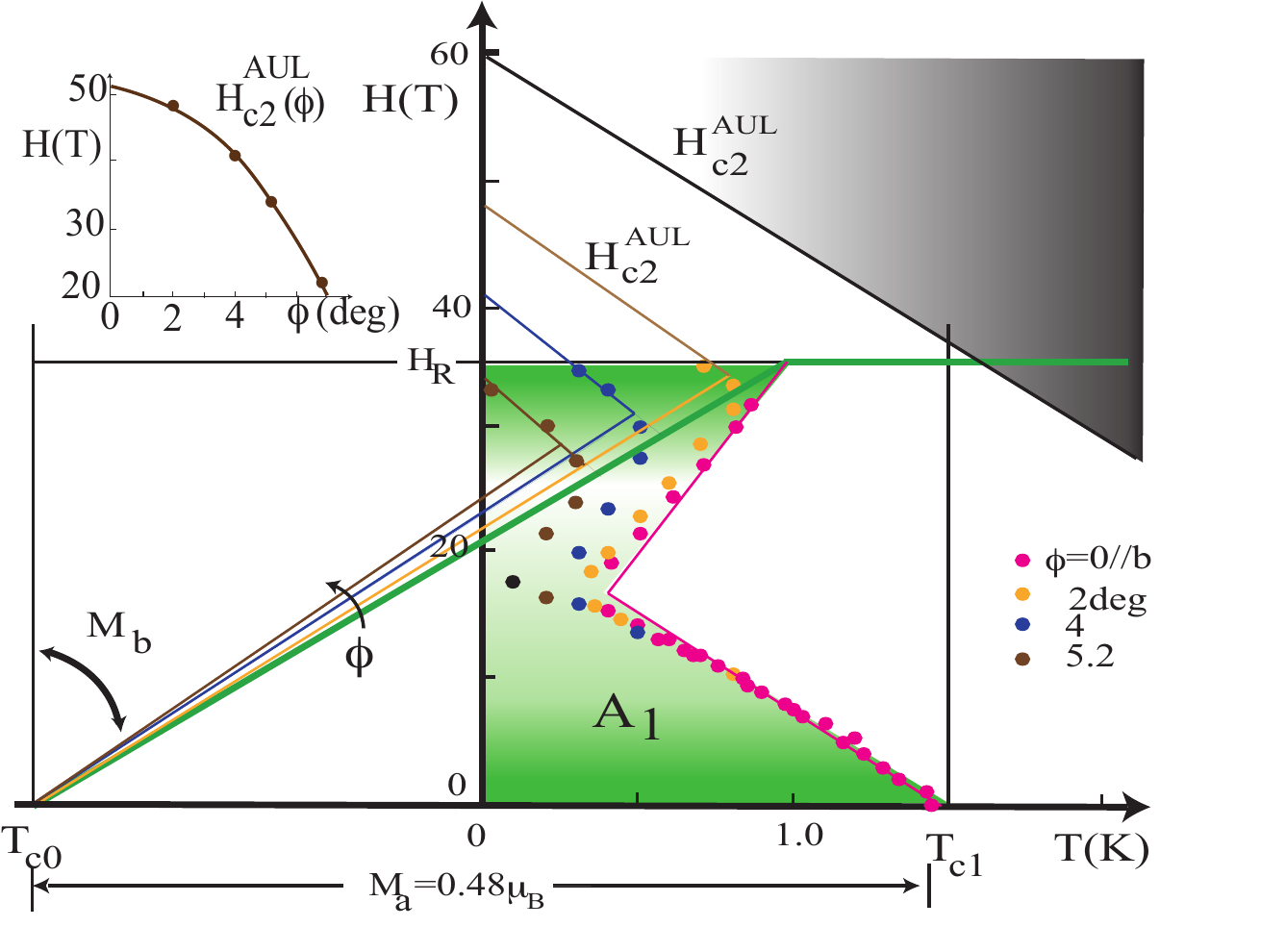}
\caption{(color online) 
$H_{\rm c2}(\phi)$ for $\phi$=0$^{\circ}$, 2$^{\circ}$, 4$^{\circ}$, 5.2$^{\circ}$, and 6.3$^{\circ}$
 from the $b$-axis toward the $a$-axis.
$M_b(H)$ grows slowly with increasing $\phi$. $H_{\rm c2}(\phi)$ curves bent over.
Before hitting $H_R(\phi)$ which ultimately limits it, $H_{\rm c2}(\phi)$ turns around
with the negative slope because they reach their own $H^{\rm AUL}_{\rm c2}(\phi)$.
$M_b(H)$ for each $\phi$ is estimated by Eq.~(\ref{elliptic}).
The data (dots) are from [\onlinecite{knebel}].
The inset shows $H^{\rm AUL}_{\rm c2}(\phi)$
estimated by extrapolating the straight lines toward higher fields beyond $H_R(\phi)$.
}
\label{UTe2phi}
\end{figure}

When the field tilts from the $b$-axis toward the magnetic easy $a$-axis by the angle $\phi$,
the magnetization $M_b(H_b)$ growth becomes slow compared to that for the $b$-axis as shown in 
Fig.~\ref{UTe2phi}. Those counter-clock wise changes of $M_b(H_b)$ for various angle $\phi$ in Fig.~\ref{UTe2phi}
are estimated by the method explained in section IV.
Therefore, $H_{\rm c2}$ is bent upward in the upper part of their L-shaped ones
while the lower parts are hardly changed because this is mainly limited by the orbital suppression.
Those $H_{\rm c2}(\phi)$ curves for various $\phi$ values eventually reach their own $H^{\rm AUL}_{\rm c2}$
which depends on $\phi$, followed by the orbital suppression. Then $H_{\rm c2}(\phi)$ finally disappears abruptly
by hitting $H_R(\phi)$. If those $H_{\rm c2}(\phi)$ curves extrapolate naively to higher fields beyond $H_R(\phi)$,
we find $H^{\rm AUL}_{\rm c2}(\phi)$ as shown in the inset of Fig.~\ref{UTe2phi}, indicating that  
$H^{\rm AUL}_{\rm c2}(\phi)$  changes strongly within a few degrees, peaking at $H\parallel b$ sharply.
We are not able to explain this peaking phenomenon at this moment.
A similar peaking phenomenon is also observed in the $\theta$ side too, where the $H_{\rm c2}(\theta)$
peak occurs at $\theta\sim 35^{\circ}$.

Thus the SC region in the $\phi$-$H$ plane is quite limited to small angles up to $\phi\sim6.3^{\circ}$.
As will show next, this is similar to the $\theta$ case where the high field SC 
$H_{\rm c2}(\theta\sim 35^{\circ})$=60T is observed in a narrow angle $\theta$ region above $T_R(\theta)$.

\subsubsection{$\theta$ rotation from the $b$-axis toward the $c$-axis}

It is remarkable to see the extremely high $H_{\rm c2}(\theta=35^{\circ})\sim60$T when the field is tilted
from the $b$-axis toward the other magnetic hard $c$-axis~\cite{ran2}. This is detached 
from the low field $H_{\rm c2}(\theta)\sim8$T. This low field SC part is nearly independent of $\theta$.
This $H_{\rm c2}$ isotropy was seen also in URhGe (see Fig.~\ref{RSC}) and UCoGe.
This extremely high $H_{\rm c2}(\theta=35^{\circ})$ can be understood by the present framework as follows.

\begin{figure}
\includegraphics[width=7.5cm]{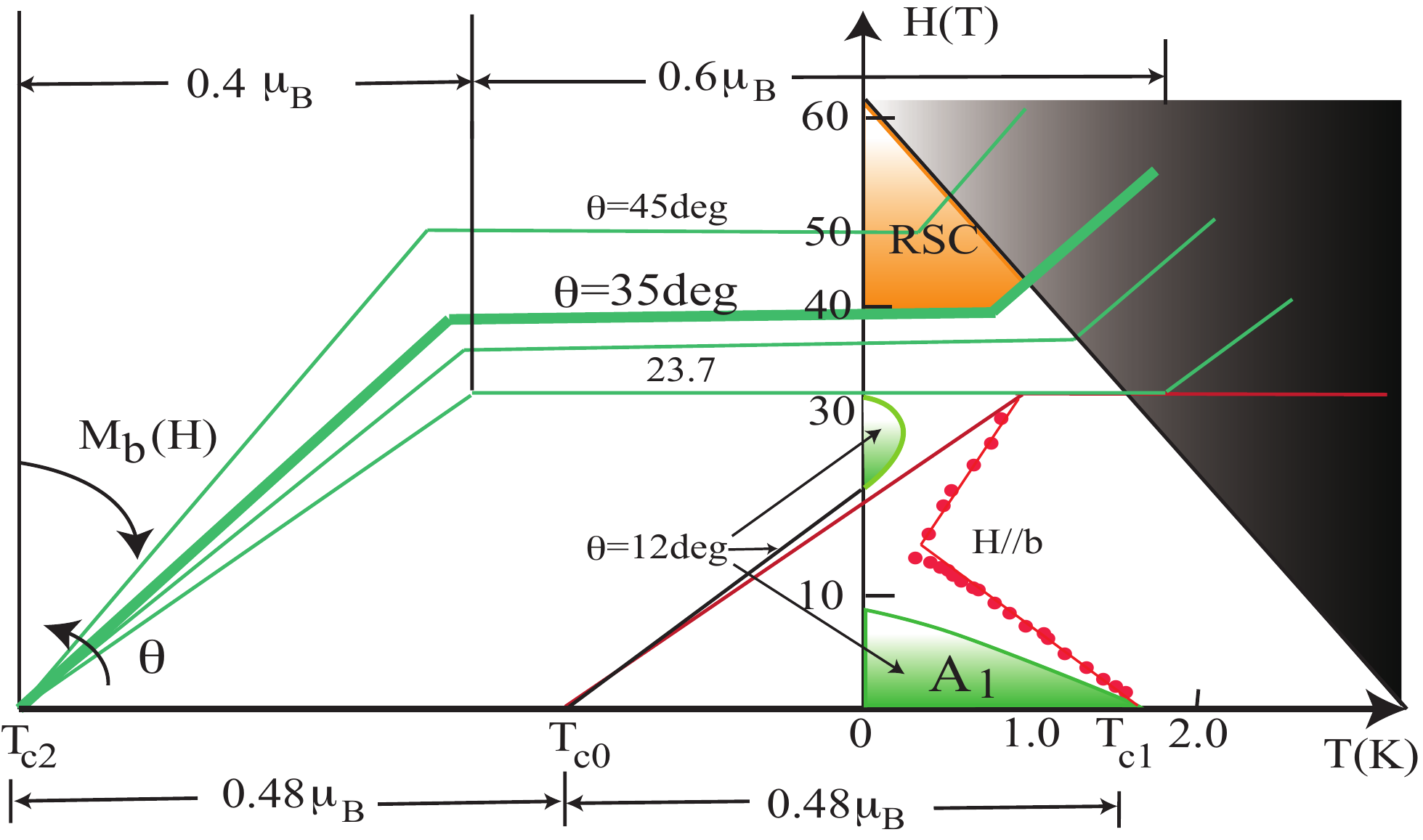}
\caption{(color online) 
$H_{\rm c2}(\theta)$ for various $\theta$, which is measured from the $b$-axis toward the $c$-axis.
The magnetization curves of $M_b(H)$ starting at $T_{c2}$ and $T_{c0}$ evaluated before (see Fig.~\ref{Mall}(b))
lead to the reentrant SC for $\theta=35^{\circ}$ in addition to the low $H_{\rm c2}$. 
For the lower angle of $\theta=12^{\circ}$ the two separate
SC are formed. Here the $\theta=0^{\circ}$ case ($H\parallel b$) is shown for reference.
It is seen that the magnetization curves only around $\theta\sim35^{\circ}$ allow RSC to appear.
}
\label{theta35}
\end{figure}

We begin with the $H\parallel b$ case discussed in Fig.~\ref{UTe2}. Upon increasing $\theta$, 
the magnetization $M_b(H)$ becomes slow to grow.
Around $\theta=12^{\circ}$ the upper part of the L-shaped $H_{\rm c2}$
separates into two parts as shown in Fig.~\ref{theta35} which is observed~\cite{georg}. And eventually this RSC part disappears
above $\theta>12^{\circ}$, leaving only the lower $H_{\rm c2}$ part at around 10T.

Further increasing $\theta$, the magnetization $M_b(H)$ starting from $T_{c2}$ becomes relevant
because as explained in Fig.~\ref{Mall}(b), $M_b(H)$ becomes small and the magnetization jump also diminishes.
Around $\theta=35^{\circ}$ the magnetization curves are just available for the reentrant SC to appear at higher fields
above the respective $H_R(\theta)$. This RSC is shown in Fig.~\ref{theta35}. 
This is because the state reached after the first order jump 
is now within the $H^{\rm AUL}_{\rm c2}$ allowed region.
Thus RSC only appears within the narrow angle region centered at $\theta=35^{\circ}$.
Those RSC regions are characterized by a triangle like shape as observed in [\onlinecite{ran2}].
This RSC shape resembles those in Figs.~\ref{URhGe1-2} and \ref{thetaURhGe} for URhGe.

\subsubsection{Phase diagrams under pressure and multiple phases}

Let us examine the pressure effects on the $H_{\rm c2}$ phase diagram, which give us another testing ground
to check the present scenario.

\begin{figure}
\includegraphics[width=8cm]{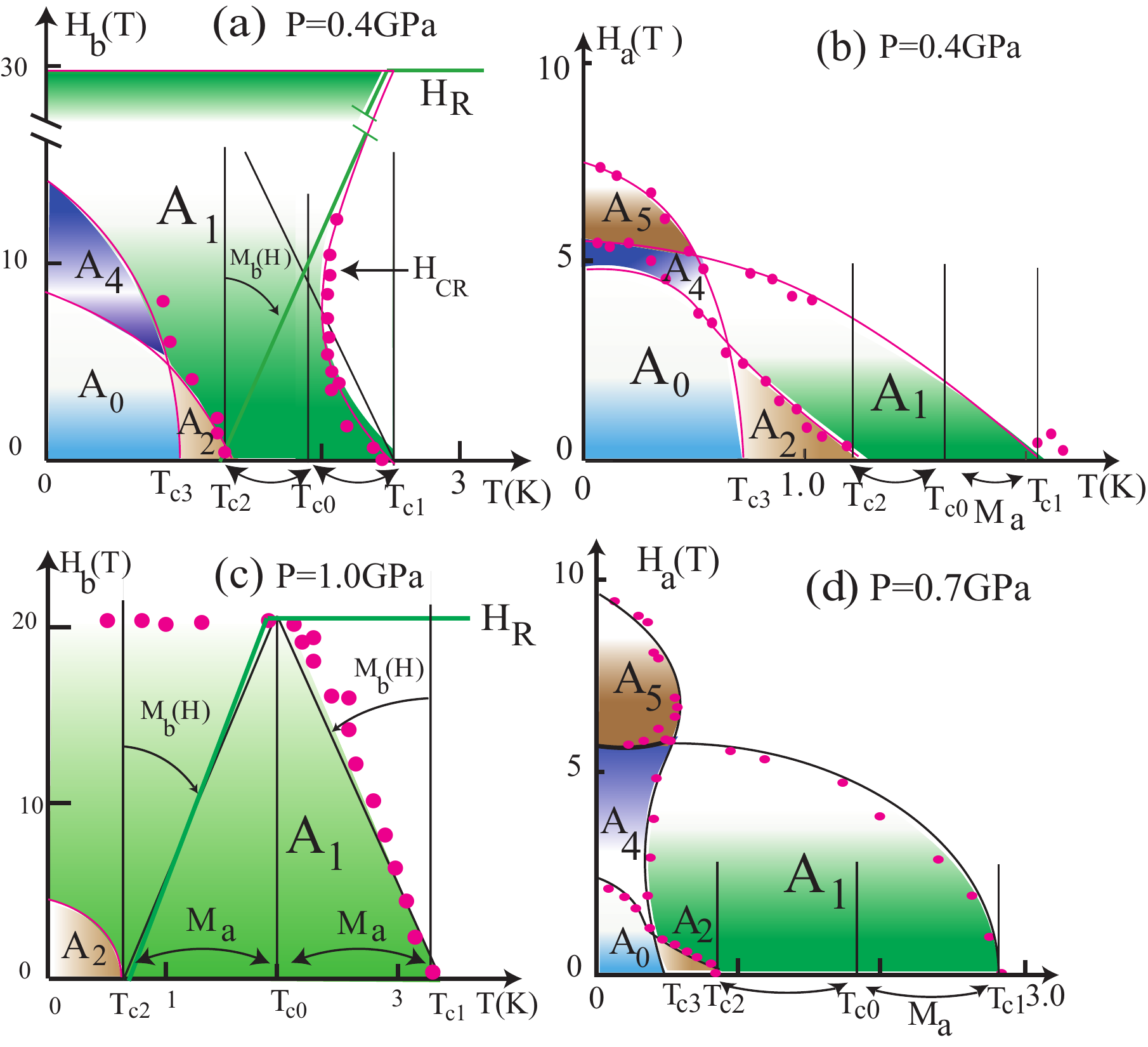}
\caption{(color online) 
$H_{\rm c2}$ and the associated internal phase transition lines under hydrostatic pressure $P$ in UTe$_2$.
(a) $P$=0.4GPa and $H//b$. (b) $P$=0.4GPa and $H//a$.
(c) $P$=1.0GPa and $H//b$. (d) $P$=0.7GPa and $H//a$. The data denoted by the red dots are from Ref.~[\onlinecite{aokiP}].
$T_{c1}$ and $T_{c2}$ at $H=0$ are split by the magnetization $M_a$ which decreases under the applied field $H_b$
as shown in (a) and (c). This decrease of $M_a(H_b)$ is compensated by growing of the magnetization $M_b(H_b)$ 
as denoted by the green lines there.
}
\label{UTe2P}
\end{figure}

In Fig.~\ref{UTe2P}~(a) we show the data (dots) of $H^b_{\rm c2}$ for $H\parallel b$ under $P=0.4$GPa~\cite{aokiP}
together with our analysis.
It is seen that since the magnetization curve $M_b(H_b)$ denoted by the green line strongly increases,
$H^b_{\rm c2}$ started at $T_{c1}$ exhibits a bent toward higher temperatures at around $H_{\rm \rm CR}$.
The two magnetization curves started from $T_{c1}$ and $T_{c2}$ meet at $H_{\rm \rm CR}$.
After passing the field $H_{\rm \rm CR}$, $H^b_{\rm c2}$ with a positive slope heads toward $H_R=30$T, 
which is observed as the first order transition~\cite{aokiP}.
The same feature is observed so far several times in URhGe under uni-axial pressure such as in Fig.~\ref{new1GPa}
and UCoGe in Fig.~\ref{UCoGe1}.

The second transition at $T_{c2}$ with the A$_2$ phase is clearly found experimentally shown there
detected by AC calorimetry by Aoki, et al~\cite{aokiP}.
Moreover, the lower $H^b_{\rm c2}$ started from $T_{c2}$ shows an anomaly at around 5T in Fig.~\ref{UTe2P}(a), 
suggesting the third transition $T_{c3}$.
This identification is quite reasonable when we see Fig.~\ref{UTe2P}(b) where the $H\parallel a$ case is displayed
for the same $P=0.4$GPa.
Indeed we can consistently identify $T_{c3}$ in this field orientation too.
According to our theory three phases $A_{1}$, $A_{2}$, and $A_{0}$ correspond to $T_{c1}$, $T_{c2}$, and $T_{c3}$ respectively
as shown there.
In the high fields, we enumerate further phases $A_{4}$ and $A_{5}$.
Those lower $T$ and high $H$ phases are the mixtures of the fundamental three phases $A_{1}$, $A_{2}$, and $A_{0}$
except for $A_{5}$, which is genuine $A_{0}$. For example, the $A_{4}$ phase consists of the $A_{1}$ and $A_{0}$ phases.

\begin{figure}
\includegraphics[width=6cm]{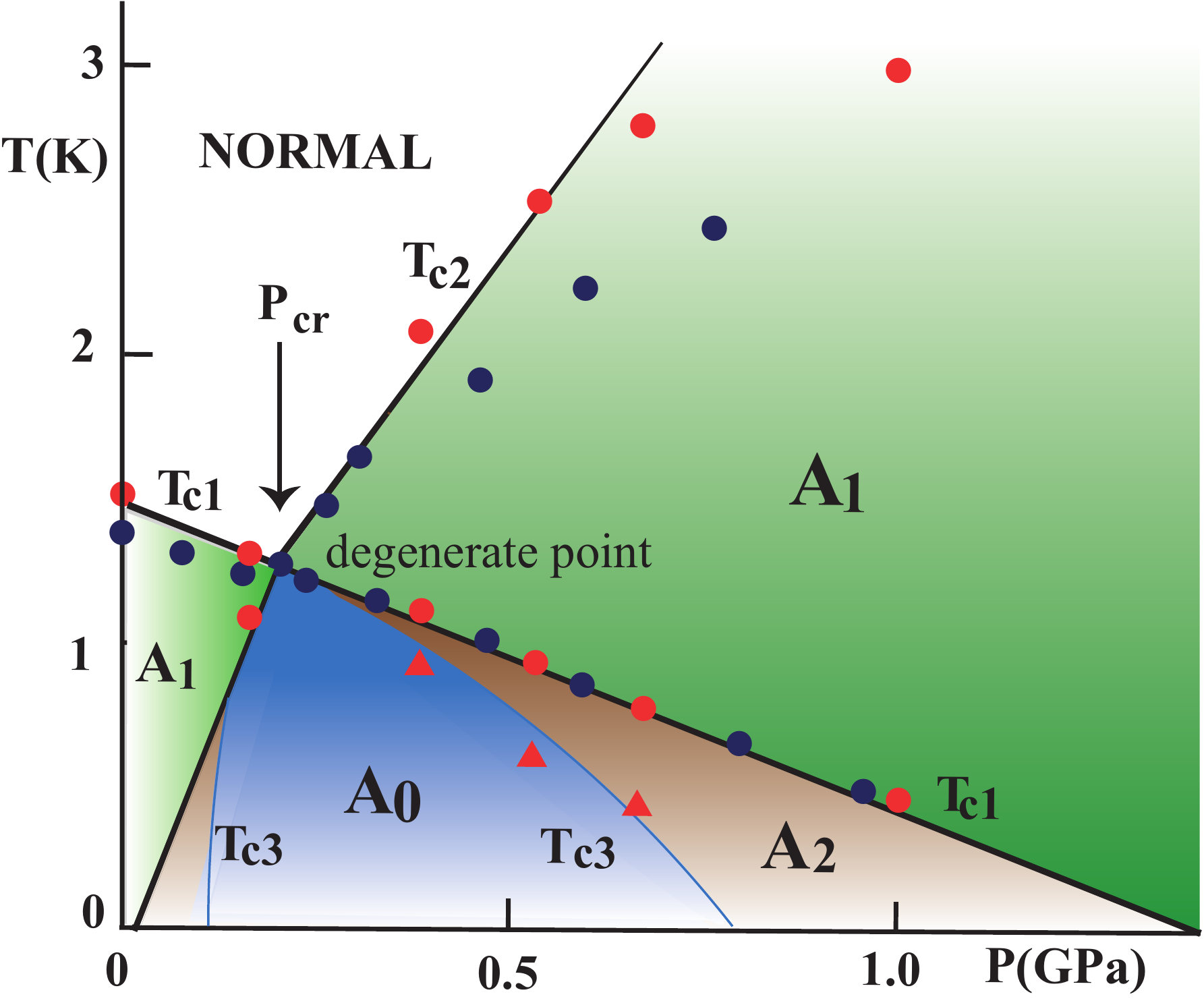}
\caption{(color online) 
$T$-$P$ phase diagram in UTe$_2$ with three transition temperatures $T_{c1}$, $T_{c2}$ and $T_{c3}$ 
corresponding to the $A_1$, $A_2$, and $A_0$ phases respectively. At the degenerate point of $P_{\rm \rm cr}=0.2$GPa 
all three phases converges. The lines for $T_{c1}$ and $T_{c2}$ as a function of $P$ indicate that the underlying
symmetry breaking field $M_a$ changes linearly with $P$, leading to the globally quadratic variation of $T_{c3}$
from the degenerate point. The red (dark blue) round dots are from the experiment 
[\onlinecite{aokiP}] ([\onlinecite{daniel}]) except for the
three red triangle points at $P$=0.40, 0.54 and 0.70GPa for $T_{c3}$, 
which are inferred from Fig.~\ref{UTe2P}.
}
\label{UTe2PT}
\end{figure}

In Fig.~\ref{UTe2P}(c) we show the data of $H^b_{\rm c2}$ for $H\parallel b$ under $P=1.0$GPa~\cite{aokiP}
together with our analysis. As $P$ increases, the first order transition field $H_R$ becomes lower, 
here it is $H_R=$20T at $P=1.0$GPa from 30T at $P=0.4$GPa.
$H^b_{\rm c2}$ just follows a straight line due to the orbital depairing all the way up to $H_R$ where the magnetization $M_b(H_b)$
denoted by the green line exhibits the magnetization jump. This jump is large enough to wipe out the SC state there.
Thus $H^b_{\rm c2}$ now follows a horizontal line at $H_R=20$T.
This is the same case as in $H^b_{\rm c2}$ seen in the ambient pressure (see Fig.~\ref{UTe2}).
The main difference from the ambient case is that the second transition at $T_{c2}$ is now visible and observable
because the FM moment $M_a$ diminishes under pressure and the pressure $P=0.4$GPa is situated
near the critical pressure at $P=0.2$GPa  (see Fig.~\ref{UTe2PT}).
This proves the consistency of our scenario.

As shown in Fig.~\ref{UTe2P}(d) where at $P$=0.7GPa for $H\parallel a$ the $H_{\rm c2}$ data points are quoted
from Ref.~[\onlinecite{aokiP}],  we draw the three continuous lines to connect those points.
We find the missing third transition along the $T$-axis at $T_{c3}$=0.5K.
Note that the tricritial point with three second order lines is thermodynamically forbidden~\cite{yip}.
The multiple phases are enumerated, such as $A_{1}$, $A_{2}$, and  $A_{0}$ at the zero-field and 
$A_{4}$, and  $A_{5}$ at finite fields. Those phases are consisting of the coexistence of the plural fundamental
three components $A_{1}$, $A_{2}$, and  $A_{0}$.
Namely, those  are characterized by A$_1$ at $T_{\rm c1}$, A$_2\rightarrow$A$_1$+A$_2$ at $T_{\rm c 2}$,
A$_0\rightarrow$ A$_1$+A$_2$+A$_0$ at $T_{\rm c3}$, A$_4\rightarrow$A$_1$+A$_0$, and A$_5\rightarrow$$A_0$.
It is understood that this phase diagram is quite exhaustive, no further state is expected
in our framework.
At the intersection points in Fig.~\ref{UTe2}(d) the four transition lines should always meet 
together according to the above general rule and thermodynamic considerations~\cite{yip}. 
The lines indicate how those three phases interact each other, by 
enhancing or suppressing. $T_{\rm c3}$ could be raised by the presence of the A$_1$ and A$_2$ phases
due to the fourth order term $Re(\eta^2_a\eta_{+}\eta_{-})$ mentioned in section II.

In Fig.~\ref{UTe2PT} we compile all the data~\cite{daniel,aokiP} of the phase transitions in the $T$-$P$ plane at $H=0$.
As $P$ increases from $P=0$, $T_{c1}$ ($T_{c2}$) decreases (increases) to meet at the critical pressure $P_{cr}=0.2$GPa
where $T_{c3}$ is also merging to converge all three transition lines. This critical pressure  corresponds to the degenerate 
point where the symmetry breaking parameter $M_a$ vanishes and the three phases $A_1$, $A_2$, and $A_0$
becomes degenerate, restoring the full SO(3) spin symmetry at this critical point.  Upon further increasing $P$,
the three phases are departing from there. The three data points for $T_{c3}$ (the three red triangles on the $T_{c3}$ line 
in Fig.~\ref{UTe2PT} are
inferred from Fig.~\ref{UTe2P}). The fact that  $T_{c1}$ and $T_{c2}$ behave linearly in $P$ is understood 
as the linear relationship between $P$ and $M_a(P)$, leading to the linear changes of $T_{c1}$ and $T_{c2}$.
This linear relationship is also seen in Fig.~\ref{URhGesigma}.
Simultaneously a strong departure of $T_{c3}$ from the critical pressure. This is because $T_{c3}$ changes
in proportion of $M_a^2$ as mentioned before (see Eq.~(\ref{tc3})).
This $T$-$P$ phase diagram is similar to that shown in Fig.~\ref{ProtoA0} globally and topologically, proving that 
the present scenario is valid for this compound too.

\section{Pairing symmetry}

\subsection{Gap symmetries and nodal structures}

The classification of the gap or orbital symmetries allowed in the present orthorhombic 
crystal has been done before~\cite{ohmi,annett}.
Among those classified pairing states, the appropriate gap function $\phi(k)$ is selected
as follows: $\phi(k)=k_ak_bk_c$ (A$_{1u}$), $\phi(k)=k_b$ (B$_{1u}$), 
$\phi(k)=k_c$ (B$_{2u}$), and $\phi(k)=k_a$ (B$_{3u}$). 
The gap structure is characterized by the line nodes for those states.
They are all candidates for URhGe and UCoGe as tabulated in Table II. This leads to the overall pairing function:
${\bf d}(k)=(\vec{a}\pm i\vec{b})\phi(k)$, which breaks the time reversal symmetry. 
This gap structure with the line nodes is consistent with the NMR experiment~\cite{manago1},
reporting that 1/T$_1$ is proportional to $T^3$ at low temperatures.
The line nodes also suggested by other experiments on UCoGe~\cite{taupin,wu}.

As for UTe$_2$, the specific heat experiments~\cite{ran,aoki2,metz,kittaka} exhibit  $C/T\sim T^2$, 
suggesting that the gap structure is characterized by
point nodes. This is also consistent with the microwave measurements~\cite{1}. 
Then we have to resort, an ad hoc orbital function, namely $\phi(k)=k_b+ik_c$ beyond the group-theoretical
classification scheme~\cite{ozaki1,ozaki2}, thus
the resulting overall pairing function is given by ${\bf d}(k)=(\vec{b}\pm i\vec{c})(k_b+ik_c)$.
This pairing state is also the time reversal broken state both in spin and orbital parts.
The point nodes are oriented along the $a$-axis determined by angle resolved specific heat experiment~\cite{kittaka}. 
This is characterized by the Weyl nodes analogous to superfluid $^3$He-A phase~\cite{mizushima1,mizushima2}.
This double chiral state both in the spin space and orbital space might be energetically advantageous
because the spin and orbital moments for Cooper pairs are parallel, namely the orbital angular moment ${\bf L}$ that is 
spontaneously induced by this
chiral state can gain the extra energy through the coupling ${\bf M}_s\cdot{\bf L}$ 
with the spontaneous magnetic moment ${\bf M}_s\propto {\bf d}\times {\bf d}^{\ast}$. 
This is consistent with the experiments by angle-resolved specific heat
measurement~\cite{kittaka}, the STM observation~\cite{mad}, and the polar Kerr experiment~\cite{hayes} among other thermodynamic
experiments~\cite{nakamine}.

\subsection{Residual density of states}

All the compounds exhibit more or less the residual density of states at the lowest $T$ limit in
the specific heat measurements~\cite{aokireview,kittaka}. 
This is not a dirt effect of the samples used, but it is intrinsic deeply rooted to the pairing state
identified as the A$_1$ phase.
In the A$_1$ phase the superconducting DOS has intrinsically the ``residual density of states''.
Since $T_{c1}$ with the A$_1$ phase is higher than $T_{c2}$ with the A$_2$ phase,
it is reasonable to expect the the DOS $N_{A_{1}}(0)$ in  the A$_1$ phase is larger than that in the A$_2$ phase,
that is, $$N_{A_{1}}(0)>N_{A_{2}}(0)$$
because in the Zeeman split bands, the major spin component band with larger DOS preferentially
forms the higher $T_c$ superconducting state rather than the minority band.
It is quite reasonable physically that in UTe$_2$ at the ambient pressure the observed 
``residual density of states'' corresponding to $N_{A_{2}}(0)$ is less than 50$\%$.

\begin{table}[t]
  \caption{Possible Pairing Functions}
  \label{pairing}
  \begin{tabular}{lcc}
    \hline
   Compound  & spin part  &  orbital part  $\phi(k)$\\
    \hline \hline
   URhGe& $\vec{a}\pm i\vec{b}$  & $k_ak_bk_c$(A$_{1u}$), $k_b$(B$_{1u}$), $k_c$(B$_{2u}$), $k_a$(B$_{3u}$) \\
  UCoGe&$\vec{a}\pm i\vec{b}$   & $k_ak_bk_c$(A$_{1u}$), $k_b$(B$_{1u}$), $k_c$(B$_{2u}$), $k_a$(B$_{3u}$) \\

  UTe$_2$& $\vec{b}\pm i\vec{c}$  & $k_b+ik_c$ \\
    \hline
\end{tabular}
\end{table}

 \subsection{Multiple phase diagram}
 
 Our three component spin-triplet state leads intrinsically  and naturally to a multiple phase diagram consisting of
 the A$_0$ phase at $T_{c3}$, A$_1$ at $T_{c1}$,  and A$_2$ at $T_{c2}$  as shown in 
 Fig.~\ref{ProtoA0} under non-vanishing symmetry breaking field due to the spontaneous moment.
 Depending on external conditions, such as $T$, $H$, and its direction, or pressure, etc,
 the structure of the multiple phase diagram is varied as explained.
In fact under $P$, the successive double transitions are clearly observed in UTe$_2$~\cite{daniel}
and they vary systematically in their $P$-$T$ phase diagram of Fig.~\ref{UTe2PT}. 
We see even the third transition centered around the critical pressure $P_{cr}=0.2$GPa.
At the ambient pressure on UTe$_2$ the occurrence of the second transition is debated~\cite{hayes,rosa},
including the detailed internal phase lines. But they agree upon the existence of the multiple phases.

As for UCoGe, the thermalconductivity experiment~\cite{taupin} indicates an anomaly at $T=0.2K$,
which coincides roughly with our prediction shown in Figs.~\ref{UCoGe1} and \ref{UCoGe-ac}. As a function of $H (\parallel b)$,
the thermalconductivity anomaly is detected as a sudden increase at $H\sim0.6H_{\rm c2}$ 
(see Fig. 5 in Ref.~[\onlinecite{wu1}]). Moreover, under $H$ parallel to
the easy $c$-axis, the $H_{\rm c2}$ curve in Fig.~\ref{UCoGe-ac}(b) shows an enhancement at  low $T$ indicative of the underlaying phase 
transition (see Fig. 2(b) in Ref.~[\onlinecite{wu}]). 
According to the NMR by Manago et al~\cite{manago1,manago2},  
$1/TT_1$ presents a similar $T$ behavior, such as a plateau at $\sim N(0)/2$ and then
sudden drop upon lowering $T$, as mentioned above. We propose to conduct further careful experiments to detect the
A$_1$-A$_2$ transitions in this compound.

In URhGe at the ambient pressure shown in Fig.~\ref{URhGe1-2} both low field phase and the RSC phase
belong to the A$_1$ phase. However,  under the uni-axial pressure along the $b$-axis, there is a good chance to observe the
second transition as explained in Figs.~\ref{new1GPa} and \ref{URhGeAll}.

Therefore to confirm the generic multiple phase diagram for all three compound shown in Fig.~\ref{ProtoA0} is essential to establish the
present scenario and also detect characteristics of each pairing state associated with those multiple phases.

 \subsection{Symmetry breaking mechanism}

For URhGe and UCoGe the ``static" FM transitions are firmly established, there is no doubt for the spontaneous FM moment
to be a symmetry breaking field.
Slow FM fluctuations are found in UTe$_2$~\cite{ran,miyake,tokunaga,sonier} which could be the origin of the symmetry breaking of
$T_{\rm c1}\neq T_{\rm c2}$ under the assumption that FM fluctuations 
are slow compared to the conduction electron motion.
A similar observation is made in UPt$_3$:  The fluctuating antiferromagnetism (AF)
at $T_N=5$K is detected only through the fast probe: ``nominally elastic'' neutron diffraction~\cite{aeppli,trappmann} 
and undetected through
other ``static'' probes, such as specific heat, $\mu$SR, and NMR. Thus the AF fluctuating time scale is an order of MHz or faster.
This is believed to be the origin of the double transition in UPt$_3$~\cite{UPt3,sauls}.

In UTe$_2$ it is essential and urgent to characterize the observed ferromagnetic fluctuations in more detail,
such as fluctuation time scale, or spatial correlation.
Elastic and inelastic neutron scattering experiments are ideal tools for it, which was the case in UPt$_3$.
It may be too early to discuss the pairing mechanism before confirming the non-unitary spin triplet state.
There already exists an opinion~\cite{appel} which advocates longitudinal ferromagnetic fluctuations to help stabilizing a
spin triplet state before the discoveries of those compounds. A problem of this sort is how to prove or refute it, otherwise
it is not direct evidence and remains only circumstantial one. We need firm objective  ``evidence'' for a pairing mechanism.
Theory must be verifiable.

 \subsection{Common and different features}
 
 As already seen, URhGe, UCoGe, and UTe$_2$ are viewed coherently from the unified point: the non-unitary triplet state.
 They share the common features:
 
 \noindent
 (1) The unusual $H_{\rm c2}$ curves occur for the field direction parallel to the magnetic hard $b$-axis,
 where the magnetization curve $M_b(H_b)$ exhibits the first order transition at $H_R$ for URhGe and UTe$_2$,
 corresponding to the FM moment rotation.
 
  \noindent
 (2) Under pressure they show the critical point behaviors $P_{cr}=0.2$GPa for UTe$_2$ and $\sigma_{cr}=1.2$GPa for URhGe
  at which the split $T_{c1}$ and $T_{c2}$ converges, leading to the $SO(3)$ spin symmetry for Cooper pairs.
 
  \noindent
 (3) The multiple phases, including the reentrant SC, are observed and explained in URhGe and UTe$_2$ and expected 
 to be confirmed for UCoGe.
 
  \noindent
 (4) The GL parameter $\kappa$ characterizing the strength of the symmetry breaking are tabulated in Table I,
 showing the similar values for three compounds.
 As a general tendency $\kappa$ is likely larger when the FM moment is larger
 because it is originated from the particle-hole asymmetry  of the
 density of states $N(0)$ at the Fermi level.
 
 There are different  features:
 
 \noindent
 (1) The nodal structures are points oriented along the magnetic easy $a$-axis in UTe$_2$ while lines in URhGe and UCoGe.
 
 \noindent
 (2) Under the ambient pressure, $H_{\rm c2}$ curves are seemingly different as in Fig.~\ref{URhGe1-2} for URhGe,
 Fig.~\ref{UCoGe1} for UCoGe, and Fig.~\ref{UTe2} for UTe$_2$. But it is now understood as mere differences
 in $T_{c0}$ or the FM moments as the symmetry breaker.

From this comparison, the superconductivity in UTe$_2$, URhGe and UCoGe should be understood
by the unified view point, which is more resourceful and productive than considered differently  and individually.

 \subsection{Double chiral non-unitary state in UTe$_2$}
 
 Since UTe$_2$ attracts much attention currently, it is worth summing up our thoughts on this system to challenge novel experiments.
 When combining the experimental observations of the chiral current along the wall by STM~\cite{mad}
 and the angle-resolved specific heat experiment~\cite{kittaka},  the double chiral non-unitary symmetry
 described by ${\bf d}(k)=({\hat b}+i{\hat c})(k_b+ik_c)$ is quite possible:
 This pairing state produces the chiral current at the edges of domain walls,
 consistent with the former observation. And it is consistent with the polar Kerr experiment~\cite{hayes} which shows the broken
 time reversal symmetry.
 In this pairing state the point nodes orient along the magnetic easy $a$-axis, which is supported by the
 angle-resolved specific heat experiment~\cite{kittaka}. This experiment further 
 indicates the unusual Sommerfeld coefficient $\gamma(H)$
 in the superconducting state for $H$ along the $a$-axis. The low energy quasi-particle excitations naively expected for
 the point nodes~\cite{miranovic} is absent. This lack of the nodal excitations is understood by taking into account
 that $T_c$ depends on $H$ through the magnetization. This is indeed consistent with the notion
 of the field-tuned SC developed throughout the present paper.

\section{Summary and Conclusion}

We have discussed the superconducting properties of  URhGe, UCoGe, and UTe$_2$ in detail
in terms of a non-unitary spin triplet pairing state in a unified way.
The spontaneous static ferromagnetic moment in URhGe and UCoGe,
and the slowly fluctuating instantaneous 
ferromagnetic moment in UTe$_2$ break the spin $SO(3)$ symmetry in the degenerate triplet pairing function with three components.
Those produce the various types of the $H_{\rm c2}$ curves that are observed.
The possible pairing function is described by the complex ${\bf d}$-vector, whose direction is perpendicular to
the magnetic easy axis at zero-field. Its direction changes under applied field parallel to
the magnetic hard $b$-axis  common in three compounds. 
This $\bf d$-vector rotation is driven by the induced magnetic moment under applied
fields. Thus the SC order parameter is tunable by the magnetic field in this sense,
ultimately leading to the reentrant SC in URhGe, S-shape in UCoGe,
and L-shape $H_{\rm c2}$ in UTe$_2$.

As for UTe$_2$, 
we can study a variety of topological properties, such as Weyl nodes associated with the point nodes, 
known in $^3$He A-phase~\cite{mizushima1,mizushima2}, which was difficult to access experimentally 
and remains unexplored in the superfluid $^3$He.
We can hope to see in UTe$_2$ similar exotic vortices  and  Majorana zero modes predicted in $^3$He phase~\cite{mizushima1,mizushima2,tsutsumi1,tsutsumi2}.

There are several outstanding problems to be investigated in future, such as the pairing mechanism leading to the present
non-unitary state where longitudinal spin fluctuations are plausible, but how to prove or to refute it. That is a question.
As a next step, microscopic theory and detailed calculations  are definitely needed beyond the present GL framework where the most simplified 
version is adopted in order to just illustrate the essential points.
For example, we did not seriously attempt to produce the observed $H_{\rm c2}$ curves quantitatively
because of the reasons mentioned at the beginning of section V.
Thus we only scratches its surface admittedly.
It is our hope that the present theory motivates ingenious experiments in this fruitful and flourishing research field.

\section*{Acknowledgments}  

The author is grateful for the enlightening discussions with Y. Shimizu, Y. Tokunaga, A. Miyake, T. Sakakibara, S. Nakamura, S. Kittaka,
G. Knebel, A. Huxley, and K. Ishida. He would especially like to 
thank D. Aoki for sharing data prior to publication and stimulating discussions. 
He thanks K. Suzuki for helping to prepare the figures.
This work is supported by JSPS KAKENHI, No.17K05553.

\end{document}